\documentstyle[preprint,prb,aps]{revtex}
\tightenlines
\begin{document}
\title{Ground state and excited state properties of LaMnO$_3$ from full-potential calculation}
\author{P.~Ravindran$^1$~\cite{byline}, A.~Kjekshus$^1$, H.~Fjellv{\aa}g$^1$, 
A.~Delin$^2$ and O.Eriksson$^3$}
\address{$^1$ Department of Chemistry, University of Oslo, Box 1033, Blindern,
N-0315, Oslo, Norway\\
$^2$ICTP, Strada Costiera 11, P.O.Box 586, 34100, Trieste, Italy \\
$^3$Condensed Matter Theory Group, Department of Physics, Uppsala
University, Box 530, 75121, Uppsala, Sweden\\
}
\date{\today}
\maketitle
\begin{abstract}
{ 
The ground state and excited state properties of the perovskite LaMnO$_3$, the
mother material of colossal magnetoresistance manganites, are calculated based on
the generalized-gradient-corrected relativistic full-potential method. The 
electronic structure, magnetism and energetics of
various spin configurations for LaMnO$_3$ in the ideal cubic perovskite structure and
the experimentally observed distorted orthorhombic structure are obtained. The excited 
state properties such as the optical, magneto-optical, x-ray photoemission (XPS), 
bremsstrahlung isochromat (BIS), x-ray absorption near edge structure
(XANES) spectra are calculated and found to be in excellent agreement
with available experimental results. Consistent with earlier observations
the insulating behavior can be obtained only when we take into account the structural
distortions and the correct antiferromagnetic ordering in the calculations.
The present results suggest that the correlation effect is not significant in LaMnO$_3$ and
the presence of ferromagnetic coupling within the $ab$ plane as well as the antiferromagnetic 
coupling perpendicular to this plane can be explained through the
itinerant band picture. As against earlier expectations, our calculations show that
the Mn 3$d$ $t_{2g}$ as well as the $e_{g}$ electrons are present in the whole valence band region.
In particular significantly large  amounts of $t_{2g}$ electrons are present in combination with the
$e_{g}$ electrons at the top of the valence band
against the common expectation of presence of only pure $e_{g}$ electrons.
We have calculated the hyperfine field parameters for the A-type antiferromagnetic and 
the ferromagnetic phases
of LaMnO$_3$ and compared the findings with the available experimental results. The role of 
the orthorhombic
distortion on electronic structure, magnetism and optical anisotropy are analyzed.
}
\end{abstract}
\section{INTRODUCTION}
\label{sec:intro}
Even though the Mn containing oxides with the perovskite-like structure have been 
studied for more than a
half century,\cite{jonker50,wollan55} various phase transitions occurring
on doping in these materials are not fully understood. 
In particular, LaMnO$_3$ exhibit rich and interesting physical properties because of the strong interplay
between lattice distortions, transport properties and magnetic ordering. This compound also have a very
rich phase diagram depending on the doping concentration, temperature and pressure; being either
antiferromagnetic (AF) insulator, ferromagnetic (F) metal or charge ordered (CO) 
insulator.\cite{schiffer95}
The magnetic behavior of the LaMnO$_3$ perovskite is particularly interesting, because
the Jahn-Teller (JT) distortion is accompanied by the so-called A-type antiferromagnetic (A-AF) spin 
(moment) and
C-type orbital ordering (OO), i.e, alternative occupation of $d_{x^{2}-r{^2}}$ and $d_{y^{2}-r^{2}}$ in the 
$ab$ plane and same type of orbital occupation perpendicular to the $ab$ plane.\cite{mizokawa99}
Recently manganites have also been subjected to strong interest
due to their exhibition of negative colossal magnetoresistance (CMR) effects.\cite{cmr}
In particular the perovskite-oxide system La$_{1-x}$AE$_x$MnO$_3$, where $AE$ is a divalent alkali element
such as Ca or Sr, have attracted much attention primarily due to the discovery
of a negative CMR effect around the ferromagnetic transition temperature
$T_{C}$, which is located near room temperature.\cite{cmr}
The mutual coupling among the charge, spin, orbital and lattice degrees of freedom in
perovskite-type manganites creates versatile intriguing phenomena such as CMR,\cite{cmr} 
field-melting of the CO and/or
OO state(s) accompanying a huge change in resistivity,\cite{order} field-induced
structural transitions even near room temperature,\cite{fstr} field control of inter-grain
or inter-plane tunneling of highly spin-polarized carriers,\cite{field} etc.
Several mechanisms have been proposed for CMR, such as double exchange,\cite{double}
dynamical JT effect,\cite{jt} antiferromagnetic fluctuation,\cite{af}
etc., but no consensus has been attained so far about the importance of those
mechanisms. Since the spin, charge, orbital and structural ordering phenomena
may affect CMR at least indirectly it is important to obtain a full understanding
of the mechanism stabilizing the observed A-AF order in the undoped insulating mother
compound LaMnO$_3$.
\par
It has been suggested that an understanding of hole-doped LaMnO$_3$ must include, in addition
to the double-exchange mechanism,\cite{double} strong electron correlations,\cite{tokura94}
a strong electron-phonon interaction\cite{millis97} and cooperative JT distortions
associated with Mn$^{3+}$.  Several theoretical studies have been made on this material using 
the mean-field approximation\cite{kugel73,maezono98}, 
numerical diagonalization,\cite{koshibae97} Gutzwiller technique,\cite{roder96}
slave-fermion theory,\cite{ishihara97}, dynamical mean-field theory\cite{millis95,brito98},
perturbation theory\cite{takahashi98} and quantum Monte-Carlo technique.\cite{motome99}
Nevertheless it is still controversial as to what is the driving mechanism of  
the experimentally established properties, particularly the strongly incoherent charge dynamics, and what
the realistic parameters of theoretical models are.
By calculating and comparing various experimentally observed quantities one can get an idea
about the role of electron correlations and other influences on the CMR effect in these materials.
Hence, theoretical investigations of ground state and excited state properties are important
to understand the exotic physical properties of these materials. 
\par
The importance of spin and lattice degrees of freedom on the metal-insulator transition in LaMnO$_3$
is studied extensively.\cite{mryasov97}
Popovic and Satpathy\cite{popovic00} showed how the cooperative JT coupling
between the individual MnO$_6$ centers in the crystal leads to simultaneous ordering
of the distorted octahedron and the electronic orbitals.
It is now accepted that OO and magnetic ordering (MO) are closely 
correlated and that the anisotropy in the magnetic coupling originates from  
OO.\cite{igor98} So in order to understand the origin of OO,
it is important to study the energetics of different spin-ordered states.
Ahn and Millis\cite{ahn00} calculated the optical conductivity of LaMnO$_3$ using a 
tight-binding parameterization of the band structure. They noted a troubling discrepancy
with LSDA band-theory calculations\cite{solovyev96} of the optical conductivity and 
concluded with the need
for further work to find the origin of the difference. Hence, accurate calculations
of optical properties is expected to give more insight into the origin of the discrepancy.
An appreciable Faraday rotation has been observed in hole-doped LaMnO$_3$ thin 
films\cite{lawler94} and hence it has been suggested that these ferromagnetic films may
be used to image vortices in high-temperature superconductors. Further, due to the 
half-metallic behavior in the F phase of LaMnO$_3$ one can expect a large magneto-optical
effect. For this reason, we have calculated the optical and magneto-optical properties
of this material.
\par
The simultaneous presence of strong electron-electron interactions within the
transition-metal 3$d$ manifold and a sizable hopping interaction between transition metal
($T$ = Ti$-$Cu) 
3$d$ and O 2$p$ states are primarily responsible for the wide range of properties
exhibited by transition-metal oxides. Often the presence of a strong intraatomic
Coulomb interaction makes a single-particle description of such systems
inadequate. Due to this deficiency, the density-functional calculations often
fail\cite{pari95,yang00} to predict the insulating behavior of LaMnO$_3$.
To correct this deficiency of the local spin-density approximation (LSDA)  
to give the right insulating
properties of the perovskites, LSDA+$U$ theory\cite{anisimov91,solovyev94}
is applied, where $U$ is the intrasite Coulomb repulsion. 
From experimental and theoretical studies it has been believed that the electron correlations
in the La$T$O$_3$ ($T$ = Ti$-$Cu) series are very important and should be considered more
rigorously beyond LSDA. Hence, LDA+$U$ is 
applied\cite{solovyev96} to obtain magnetic moments and fundamental gap in good
agreement with experiment. 
The calculations for La$T$O$_3$ by Solovyev {\em et al.}
\cite{solovyev96} showed
that the correlation correction was significant for Ti,V,Co but less important for Mn.
However, in this study the calculated intensity of the optical conductivity
were found to be much smaller than the experimental results in the whole
energy range. 
Hu {\em et al.}\cite{hu00} reported that to get the correct experimental ground state 
for LaMnO$_3$, it is necessary to take JT distortion, 
electron-electron correlation and AF ordering simultaneously into consideration.
From the observation of large on-site Coulomb $U$ and exchange $J$, obtained from 
"constrained LDA calculations", Satpathy {\em et al.}\cite{satpathy96} indicated the importance of
correlation effects in LaMnO$_3$.
Held and Vollhardt,\cite{held00} using the dynamical mean-field theory, emphasize the
importance of electronic correlations from the local Coulomb repulsion for understanding
the properties of manganites.
Maezono {\em et al.}\cite{maezono98} pointed out that the electron correlations remain
strong even in the metallic state of doped manganites.
Photoemission studies on doped manganites gave an electron-electron correlation effect
with $U$ = 7.5\,eV for the Coulomb repulsion.\cite{inoue95,bocquet92}
\par
In contrast to the above observations, it has been shown that many
aspects of the ground state as well as single-electron excited state properties of
LaMnO$_3$ and related compounds can be described satisfactorily in terms of LSDA energy
bands.\cite{sarma96,pickett96,solovyevprl96}
Density functional calculations
also show strong couplings between lattice distortions, magnetic order, and electronic
properties of LaMnO$_3$. In particular, it is found that without lattice distortions 
LaMnO$_3$ would have a F metallic ground state, and even if forced to be A-AF, it
would still be metallic.\cite{pickett96} Further, Sarma {\em et al.}\cite{sarma96} indicated that the 
electron-electron correlation is unimportant due to a relatively large hopping parameter $t$
and a large screening effect.
The LSDA studies\cite{satpathy96,pickett96,solovyevprl96,sarma95,singh98}
show that substantially hybridized bands derived from majority-spin Mn $e_{g}$ 
and O $p$ states dominate the electronic structure near the Fermi energy ($E_{F}$). X-ray absorption 
spectroscopy (XAS) data show that several apparent peaks exist up to 5\,eV above $E_{F}$, 
but some disputes still remain about the origins of these conduction-band peaks.\cite{abbate92,park96,saitoh95}
The reported positions of the $e_{g}^{1\uparrow}$ bands in LaMnO$_3$  
differ.\cite{satpathy96,saitoh95}  
The above mentioned studies indicate that it is important to study the significance of correlation effects
in LaMnO$_3$.
\par
Some of the features lacking in most of the theoretical studies on LaMnO$_3$ originate
from the fact that
they have often resulted from the use of the atomic-sphere approximation (ASA), i.e. the calculations 
have not included the nonspherical part of the potential and also used a minimal
basis set. 
Further the cubic perovskite structure is frequently assumed\cite{pari95,bouarab96} and 
the significant structural distortions are not taken into account. Apart from these, owing to the presence of
magnetic ordering, relativistic effects such as spin-orbit coupling
may be of significance in this material, which was not been included in earlier studies.
Moreover, it is shown that instead of using the uniform electron gas limit for exchange
and correlations (corresponding to LSDA) one can improve the outcome by including the 
inhomogeneity effects through
the generalized-gradient approximation (GGA).\cite{sawada97}
To overcome the above mentioned deficiencies we have used a generalized-gradient-corrected, relativistic
full-potential method with the experimentally observed orthorhombic distorted perovskite structure
as the input in the present calculation.

\par
The rest of the paper is organized as follows. The structural aspects and the computational
details about the calculations of the electronic structure, energetics of different magnetic
phases, optical properties, magneto-optical properties, XPS and XANES features are given in Sec.\,\ref{sec:details}.
In Sec.\,\ref{sec:resdis} we give the orbital, angular momentum and site-projected density of states (DOS)
for LaMnO$_3$ in the ground state, the spin-projected DOS for various magnetic state. The
calculated band structure for the A-AF and F phases of the distorted 
orthorhombic structure are given. The role of the structural distortion
on the electronic structure, optical and magnetic properties is analyzed. Calculated
magnetic properties are compared with available experimental and theoretical studies.
The origin of excited state properties such as XPS, BIS, XANES, optical properties and magneto-optical
properties is analyzed through the electronic structure and compared with
available experimental spectra. Finally in Sec.\,\ref{sec:con} we have summarized the important 
findings of the present study.

\section{Structural aspects and computational details}
\label{sec:details}
\subsection{Crystal and magnetic structure}
\label{ssec:str}
LaMnO$_3$ is stabilizes in the orthorhombic GdFeO$_3$-type structure\cite{matsumoto70,elemans71} 
(comprising four formula units; space group $Pnma$) as shown in Fig.\,\ref{fig:str}.
It can be viewed as a highly
distorted cubic perovskite structure with a quadrupled unit cell ($a_{p}\sqrt{2}$,
$a_{p}\sqrt{2}$, 2$a_{p}$
where $a_{p}$ is the lattice parameter of the cubic perovskite structure.
The structural parameters used in the calculations are  taken from a 4.2\,K 
neutron diffraction study\cite{elemans71} with
$a$ = 5.742, $b$ = 7.668 and $c$ = 5.532\,{\AA} and atom positions:
La in 4c (0.549 0.25  0.01), Mn in 4a (0 0 0), O(1) in 4c ($-$0.014 0.25 $-$0.07)  and O(2) in 8d 
(0.309 0.039 0.244).
The electronic configuration of Mn$^{3+}$ in LaMnO$_3$ is postulated as
$t_{2g}^{3 \uparrow} e_{g}^{1 \uparrow}$ and hence, it is a typical JT system. 
Basically, two different types of distortions are included in the structure shown 
in  Fig.\,\ref{fig:str}. One is a
tilting of the MnO$_6$ octahedra around the cubic [110] axis as in
GdFeO$_3$ so that the Mn-O-Mn angle changes from 180$^{o}$ to $\sim$ 160$^o$ which is not directly 
related with the JT effect, but is
attributed to the relative sizes of the components, say, expressed in terms of the tolerance factor
$t_{p} = \frac{R_{La} + R_{O}}{\sqrt{2}(R_{Mn}+R_{O})}$, where $R_{La}$, $R_{Mn}$ and R$_O$
are the ionic radii for La, Mn and O respectively giving $t_{p}$ = 0.947 for LaMnO$_3$.
The rotation of the MnO$_6$ octahedra facilitates a more efficient space filling. The second type of
crystal distortion in LaMnO$_3$ is the deformation of the MnO$_6$ octahedra caused by the 
JT effect, viz. originating from orbital degeneracy. This may be looked upon as a cooperative 
shifting of the oxygens within the 
$ab$ plane away from one of its two nearest neighboring Mn atoms towards the others,
thus creating long and short Mn-O bond lengths (modified from 1.97\,{\AA} for cubic case to 1.91,1.96
and 2.18\,{\AA} for the orthorhombic variant) perpendicularly arranged with respect to the Mn atoms. 
The long bonds can be
regarded as rotated
90$^{o}$  within $ab$ on going from one Mn to the neighboring Mn.\cite{goodenough55}
The structural optimization\cite{sawada97} wrongly predicts that the ground state of LaMnO$_3$
is close to that of cubic perovskite with the F state lower in energy
than the experimentally observed AF state. Hence it is important to study the relative
stability between the orthorhombic and cubic phases for different magnetic
arrangements. Consequently we have made calculations both for the orthorhombic ($Pnma$) 
as well as the ideal cubic perovskite variants.
For the calculations of the undistorted cubic variant we have used the experimental 
equilibrium volume for the orthorhombic structure.

\par
When
LaMnO$_3$ is in the AF state there are three possible magnetic arrangements 
according to interplane
and intraplane couplings within (001) plane. (i) With interplane AF coupling and 
intraplane F coupling
the A-AF structure arises. (ii) The opposite structure of A-AF, where the 
interplane coupling is F, but the intraplane coupling is AF is called C-AF
structure. 
In the C-type cell all atoms have two F and four AF nearest neighbors
whereas the reverse is true for A-type AF.
(iii) If both the inter- and intraplane couplings are AF the G-AF
structure should arise.\cite{bouarab96} 
In the G-type AF lattice, each Mn atom is surrounded by six Mn neighbors whose
spins are antiparallel to the chosen central atom. 
Among the several possible magnetic orderings, 
the experimental studies
show that for LaMnO$_3$ the A-AF ordering is the ground state with a N\'{e}el temperature of 140\,K.
In the cubic case we have made calculations for the AF structures
in the following way. The A-AF structure contains two formula units, and is a tetragonal
lattice with $a$ = $b$ = $a_{p}$ and $c$ = 2$a_{p}$. 
The C-AF structure is also tetragonal with $a$ = $b$ = $\sqrt{2}a_p$
and $c$ = a$_p$. Finally, the G-AF structure is a face centered cubic structure with
$a$ = $b$ = $c$ = 2$a_p$ with eight formula units, where two nonequivalent 3$d$ metals with spins up
and down replace Na and Cl in an NaCl configuration.  This magnetic structure can be 
viewed as consisting of two 
interpenetrating face-centered lattices with opposite spin orientation.

\subsection{Computation details for the LAPW calculations}
These investigations are based on {\em ab initio} electronic structure calculations
derived from spin-polarized density-functional theory (DFT). For the XANES, orbital-projected 
DOS and the screened plasma frequency calculations we have
applied the full-potential linearized-augmented plane wave (FPLAPW) method\cite{wien} 
in a scalar-relativistic version, i.e. without spin-orbit (SO) coupling. In the calculation we 
have used the atomic sphere radii ($R_{MT}$) of 2.2, 2.0 and 1.5 a.u. for La, Mn 
and O, respectively.  Since the spin
densities are well confined within a radius of about 1.5 a.u., the resulting
magnetic moments do not depend strongly on the variation of the atomic sphere radii.
The charge density and
the potentials are expanded into lattice harmonics up to $L$ = 6 inside the spheres and
into a Fourier series in the interstitial region. 
The initial basis set included 4$s$, 4$p$, 3$d$ functions and 3$s$, 3$p$ semicore
functions at the Mn site, 6$s$, 6$p$, 5$d$ valence and 5$s$, 5$p$ semicore functions for the La site
and 2$s$, 2$p$ and 3$d$ functions for the O site. This set of basis functions was supplemented
with local orbitals for additional flexibility in representing  
the semicore states and for relaxing the linearization errors generally. 
Due to the linearization errors DOS are reliable to about 1$-$2\,Ry above the Fermi
level. So, after selfconsistency was achieved for this
basis set we included a few high-energy local orbitals; 6$d$-, 4$f$-like function for La,
5$s$-, 5$p$-like functions for Mn and 3$p$-like functions for O atoms. 
The effects of exchange and correlation are treated within the
generalized-gradient-corrected local spin-density approximation using the
parameterization scheme of Perdew {\em et al.}\cite{pw96} To ensure 
convergence for the Brillouin
zone (BZ) integration 243\,{\bf k} points in the irreducible wedge of first 
BZ were used.
Self-consistency was achieved by demanding the convergence of the total
energy to be smaller than 10$^{-5}$\,Ry/cell. This corresponds to a convergence of the
charge below 10$^{-4}$ electrons/atom. 

\subsection{Computational details for the FPLMTO calculations}
The full-potential LMTO calculations\cite{wills} presented in this paper
are all electron, and no shape approximation to the charge density or
potential has been used. The base geometry in this computational method
consists of a muffin-tin part and an interstitial part.
The basis set is comprised of augmented linear muffin-tin
orbitals.\cite{oka75} Inside the muffin-tin
spheres the basis functions, charge density and potential are
expanded in symmetry-adapted spherical harmonic functions together
with a radial function and a Fourier series in the interstitial. In
the present calculations the spherical-harmonic expansion of the charge
density, potential and basis functions were carried out up to
$L$ = 6. The tails of the basis functions outside their
parent spheres are linear combinations of Hankel or Neumann functions
depending on the sign of the  kinetic energy of the basis function in
the interstitial region. For the core charge density, the Dirac equation
is solved self-consistently, i.e. no frozen core approximation is used. The
calculations are based on the generalized-gradient-corrected-density-functional 
theory as proposed by Perdew {\em et al.}\cite{pw96}

\par
The SO term is included directly in the Hamiltonian matrix elements
for the part inside the muffin-tin spheres,
thus doubling the size of the secular matrix for a spin-polarized calculation.
Even though the La $4f$ states are well above $E_{F}$ their contribution to the
magnetic and structural properties are very important.\cite{sawada97} So we have
included these orbitals in all our calculations.
Moreover, the present calculations make use of a so-called
multibasis, to ensure a well converged wave function. This means that
we use different Hankel or Neuman functions each attaching to its own
radial function.
We thus have two $6s$, two $5p$, two 6$p$, two 5$d$ and two 4$f$ orbitals
for La, two $4s$, two $5p$ and three 3$d$ orbitals for Mn, two $2s$, three
$2p$ and two $3d$ orbitals for O in our expansion of the wave
function. In the method used here, bases corresponding to multiple principal
quantum numbers are contained within a single, fully hybridizing basis set.
The direction of the moment is chosen to be (001).
The calculations were performed for the cubic perovskite structure as
well as the orthorhombic GdFeO$_3$-type structure in
nonmagnetic (P), F, A-AF, C-AF and G-AF states.
The {\bf k}-space integration was performed using the special point
method with 284 {\bf k} points in the irreducible part of 
first BZ for the orthorhombic structure and the same density of {\bf k} points
were used for the cubic structure in the actual cell as well as the supercell.
All calculations
were performed using the experimental structural parameters 
mentioned in Sec.\,\ref{ssec:str} for both the nonmagnetic and 
spin-polarized cases. 
Using the self-consistent potentials obtained from our calculations,
the imaginary part of the optical dielectric tensor and
the band structure of LaMnO$_3$ were calculated for the A-AF and 
F cases. The density of states were calculated for 
the P, F, A-AF, C-AF and G-AF phases in the cubic as well as
orthorhombic structure using the linear tetrahedron technique.

\subsection{Calculation of optical properties}
Once energies $\epsilon_{{\bf k} n}$ and functions $|{\bf k}n\rangle$ for the $n$ bands are obtained self
consistently, the 
interband contribution to the imaginary part of the dielectric functions
$\epsilon_{2}$($\omega$) can be calculated by summing transitions from occupied to
unoccupied states (with fixed {\bf k} vector) over BZ, weighted
with the appropriate matrix element for the probability of the transition.
To be specific, the components of $\epsilon_{2}$($\omega$) are given by
\begin{eqnarray}
\epsilon_{2}^{ij}(\omega)  & = & \frac{Ve^2}{2\pi\hbar m^2 \omega^2}
\int d^3\! k
\sum_{ n n^{\prime} }
\bigl\langle {\bf k} n \big |  p_{i} \big | {\bf k}
n^{\prime} \bigr\rangle
\bigl\langle {\bf k} n^{\prime} \big |  p_{j} \big | {\bf
k}
n \bigr\rangle \times  \nonumber \\
\label{e2}
& & f_{{\bf k}n}\, \bigl(1 - f_{{\bf k} n^{\prime}}\bigr)
\delta\bigl( \epsilon_{{\bf k} n^{\prime}} - \epsilon_{{\bf k} n} - \hbar
\omega
\bigr) ,
\end{eqnarray}
Here ($p_{x}$,$p_{y}$,$p_{z}$) = {\bf p} is the momentum operator and $f_{{\bf k}n}$
is the Fermi distribution.
The evaluation of the matrix elements in Eq. (\ref{e2}) is done over
the muffin-tin and interstitial regions separately. Further details about the
evaluation of matrix elements are given elsewhere.\cite{alouani}
Due to the orthorhombic structure
of LaMnO$_3$ the dielectric function is a tensor. By an appropriate choice
of the principal axes we can diagonalize it and restrict our considerations
to the diagonal matrix elements. We have calculated the three components 
$E\|a$, $E\|b$ and $E\|c$ of the dielectric
constants corresponding
to the electric field parallel to the crystallographic axes
$a$, $b$ and $c$, respectively.
The real part of the
components of the dielectric tensor $\epsilon_{1}$($\omega$) is
then calculated using the Kramer-Kronig transformation. The knowledge of
both the real and
imaginary parts of the dielectric tensor allows the calculation of
important optical constants. In this paper, we present 
the reflectivity $R(\omega$), the absorption coefficient $I(\omega$),
the electron energy loss spectrum $L(\omega$), as well as the refractive
index $n$ and the extinction coefficient $k$.
The calculations yield unbroadened functions. To reproduce 
the experimental conditions more correctly,
it is necessary to broaden the calculated spectra. The exact form of the
broadening function is unknown, although comparison with measurements
suggests that the broadening usually increases with increasing excitation
energy. Also the instrumental
resolution smears out many fine features.
To simulate these effects the
lifetime broadening was simulated by convoluting the absorptive part of the
dielectric
function with a Lorentzian, whose full width at half maximum (FWHM) is equal to
$0.01(\hbar\omega)^2$\,eV. The experimental resolution was simulated by
broadening the
final spectra with a Gaussian, where FWHM is equal to 0.02 eV.

\subsection{Calculation of magneto-optical properties}
The magneto-optic effect can be described by the off-diagonal elements of the dielectric
tensor which originate from optical transitions with a different frequency dependence
of right and left circularly polarized light because of SO splitting of the
states involved. For the polar
geometry, the Kerr-rotation ($\theta_{K}$) and ellipticity ($\eta_{K}$) are related to the optical
conductivity through the following relation:
\begin{eqnarray}
\label{eq:kerr1}
\frac{1+\tan(\eta_{K})}{1-\tan(\eta_{K})} e^{2i\theta_{K}}
= \frac{(1+n_{+})(1-n_{-})}{(1-n_{+})(1+n_{-})},
\end{eqnarray}
where $n_{\pm}^{2}$ 
in terms of conductivities are
\begin{eqnarray}
n^2_{\pm} = 1 + \frac {4\pi i}{\omega}(\sigma_{xx} \pm i\sigma_{xy}).
\end{eqnarray}
For small Kerr-angles, Eq.~(\ref{eq:kerr1}) can be simplified to\cite{kahn69}
\begin{eqnarray}
\theta_{K}+i\eta_{K} = \frac{-\sigma_{xy}}{\sigma_{xx}\sqrt{1+\frac{4\pi
i}{\omega}\sigma_{xx}}}.
\label{eqn:kerr}
\end{eqnarray}
The magnetic circular birefringence, also called the Faraday
rotation $\theta_{\rm F}$, describes the rotation of the polarization plane of
linearly polarized light on transmission through matter magnetized in the
direction of the light propagation.
Similarly, the Faraday ellipticity $\eta_{F}$, which is also known as the
magnetic circular dichroism, is proportional to the difference of the
absorption for right- and left-handed circularly polarized light.\cite{faraday}
Thus, these quantities are simply given by\cite{reim90}
\begin{eqnarray}
\theta_{\rm F} + i\eta_{\rm F}=
\frac{\omega d}{2c}\left( n_+ - n_- \right),
\end{eqnarray}
where $c$ is the velocity of light in vacuum, and $d$ is the thickness of
the thin film.

\subsection{Calculation of XPS and BIS spectra}
Within the so-called single-scatterer final-state approximation\cite{winter84,redinger86}
(free propagation of the photoelectrons through the crystal, loss of {\bf k}-dependent
information and neglect of surface effects)
the photocurrent is a sum of local (atomic like), partial ($\ell$-like)
DOS weighted by cross sections (transition probabilities).
As the theoretical framework of the XPS intensity calculations is given 
earlier,\cite{marksteiner86} only a brief description of the main points is outlined here.
For high incident energies of XPS ($\sim$ 1.5\,keV) the low-energy electron-diffraction
function\cite{redinger86} can be simplified and the fully relativistic angle-integrated intensity
$I(E,\omega)$ can be written as

\begin{eqnarray}
I(E,\omega) = \sum_{\tau} \sum_{\kappa}\sigma^{\tau}_{\kappa}(E,\omega)D^{\tau}_{\kappa}(E)
 \delta(E_{F} + \hbar\omega - E)
\end{eqnarray}
\[
\mbox{$\kappa$} = \left\{ \begin{array}{ll}
\mbox{-$\ell-1$} & \mbox{$j=\ell +\frac{1}{2}$}\\
          \mbox{$\ell$} & \mbox{$j = \ell - \frac{1}{2}$}
\end{array}
\right.  \]
where $\omega$ denotes the energy of the incident photons.
This expression has been cast into a Fermi golden-rule form, where 
$D^{\tau}_{\kappa}(E)$ are the partial DOS functions for the $\kappa^{th}$
channel at the $\tau^{th}$ site, and are obtained from full-potential LMTO calculation.
The partial, angular momentum-dependent cross sections $\sigma^{\tau}_{\kappa}(E,\omega)$ of
the species $\tau$ can 
be obtained from 

\[
\sigma^{\tau}_{\kappa}(E,\omega) 
= \sum_{\kappa^{\prime}}(2j+1) 
\left( 
\begin{array}{ccc}
j & 1 & j^{\prime}\\
\frac{1}{2} & 0 &- \frac{1}{2}\\
\end{array} 
\right) ^{2} 
\times \left[ M_{kk^{\prime}}^{\tau}(E,\omega)\right]^{2} 
\]
where

\begin{eqnarray}
M_{kk^{\prime}}^{\tau}(E,\omega) = \int_{0}^{R_{mt}} R^{\tau}_{\kappa}(r_{\tau},E)\nabla V^{\tau}
(r_{\tau})Z_{\kappa^{\prime}}^{\tau}(r_{\tau},\omega)r_{\tau}^{2}dr_{\tau}
\label{eqn:sigma}
\end{eqnarray}
In Eq.\ref{eqn:sigma} the radial functions $R_{\kappa}^{\tau}(r_{\tau},E)$ and 
$Z_{\kappa^{\prime}}^{\tau}(r_{\tau},E)$ differ conceptually only by different 
single-site scattering normalizations.
The radial gradients $\nabla V^{\tau}(r_{\tau})$ refer to the $\tau$-th scattering potential.
The Wigner 3j-symbols \[ \left( \begin{array}{ccc}
j & 1 & j^{\prime}\\
\frac{1}{2} & 0 &- \frac{1}{2}\\
\end{array}
\right)\]
automatically take care of the dipole selection rules.
The relativistic cross sections $\sigma_{\kappa}^{\tau}(E,\omega)$ are calculated
using the muffin-tin part of the potential
over the energy range $E$ of the DOS functions for the fixed incident photon energy $\omega$.
This expression has been evaluated for LaMnO$_3$
from the potentials and DOS functions by a fully relativistic full-potential LMTO 
self-consistent calculation. Because the cross sections (matrix elements)
are energy dependent, the theoretically predicted spectra will depend on the energy 
chosen for the calculation. To be consistent with the reported XPS data, we have made
all calculations with the fixed incident photon energy $\omega$ = 1253.6\,eV 
(Mg$K\alpha$ line) used in the experimental study. The finite lifetime
of the photoholes was taken into account approximately by convoluting the spectra using
a Lorentzian with a energy-dependent half-width. The FWHM of this 
Lorentzian is zero at $E_{F}$ and increases linearly with the binding energy, as 
0.2($E_{F}-E$). In addition to the Lorentzian life-time broadening, the spectra were
broadened with a Gaussian of halfwidth 0.8\,eV to account for spectrometer resolution.
The procedure we adapt to calculate XPS and BIS spectra is as follows. First,
we take the partial DOS functions ($E < E_{F}$ for XPS and $E > E_{F}$ for
BIS) from a full-potential LMTO calculations. We multiply
these by the calculated $\ell$ dependent cross-sections for all band energies and
sum them to get a total spectral-like function. Next, to get a good fit to the data
we broaden this with an energy-dependent Lorentzian function to simulate what we call
inherent life-time effects due to the coupling of the excited outgoing electron from the
crystal. This broadening is zero at $E_{F}$ and goes as the square of the energy below
$E_{F}$. Then we fit the leading edge to the experimental data with a Gaussian broadening
to simulate the instrument resolution. This broadening is same for all states. Finally
we add a background function to get the experimental profile. While the calculations were
performed at the one-electron level, we believe this procedure should capture the essence
of the photoemission process and lead to a meaningful comparison between calculated and
observed spectra.

\subsection{Calculation of XANES}
The theoretical x-ray absorption spectra for LaMnO$_3$ were computed within the dipole
approximation from the FLAPW\cite{wien} partial DOS along the lines described by 
Neckel {\em et al.}\cite{neckel75}
The intensity $I(\omega) = E-E_{c}$ arising from transitions from initial VB states (energy
$E$ and and angular momentum $\ell$) to a final core state ($E_{c}$,$\ell^{\prime}$) is given by
\begin{eqnarray}
\frac{I_{\tau n^{\prime}l^{\prime}}(\omega)}{\omega^{3}} = \sum_{\ell}
W_{\ell\ell^{\prime}} M_{\tau}^{2}(\ell,n^{\prime}\ell^{\prime},E) D_{\ell}^{\tau}(E)
\delta(E-E_{n^{\prime}\ell^{\prime}}^{core} - \hbar\omega)
\end{eqnarray}
where the matrix elements are given by
\[
M_{\tau}^{2}(\ell,n^{\prime}\ell^{\prime},E) = \frac{ \left[\int_{0}^{R_{\tau}}P_{\ell}^{\tau}(r,E)
rP_{n^{\prime}\ell^{prime}}(r)dr\right]^{2}}
{\int_{0}^{R_{\tau}}P_{\ell^{\tau}}(r,E)^{2}dr}
\]
\noindent
$n^{\prime} and \ell^{\prime}$ represent principal and angular momentum quantum numbers for
the core states. $D_{\ell}^{\tau}(E)$ is the partial DOS of atom $\tau$ with angular momentum
$\ell$, $P_{\ell}^{\tau}$ and $R_{\tau}$ are the radial wave function and atomic sphere radii of
atom $\tau$. The transition coefficient $W_{\ell\ell^{\prime}}$ can be calculated
analytically according to the following equation,
\[
W_{\ell\ell^{\prime}} = \frac{\ell + 1}{2\ell + 1} \delta_{\ell,\ell^{\prime}-1}
+ \frac{\ell}{2\ell - 1} \delta_{\ell,\ell^{\prime}+1}
\]
In the case of absorption spectra $D_{\ell}^{\tau}(E>E_{F}) = 0$. To account for instrument 
resolution and life-time broadening of both core and valence states we have broadened the
calculated spectra using the Lorentzian function with FWHM of 1\,eV.

\section{Results and discussion}
\label{sec:resdis}
\subsection{Electronic structure}
\label{electronic}
\subsubsection{Electronic structure of the cubic phase}
Let us first discuss qualitative distinctions among the electronic band structures of
LaMnO$_3$ with various magnetic configurations without considering the structural distortion.
With full cubic symmetry, $t_{2g}$ and e$_g$ orbitals are three-fold and two-fold degenerate,
respectively. The orbital-projected DOS of Mn 3$d$ electrons in the F phase of
cubic LaMnO$_3$ are shown in Fig.\,\ref{fig:cldos},
where the $t_{2g}$ states are away from $E_{F}$ and also rather narrow. However, the
$e_{g}$ levels are broadly distributed in the DOS profile. The electrons
at $E_{F}$ have both $e_{g}^{\uparrow}$ and the $t_{2g}^{\downarrow}$
electrons as shown in Fig.\,\ref{fig:cldos}. Against the pure ionic picture (where the
$e_{g}$ electrons only are expected to be closer to $E_{F}$)
there is considerable amount of $e_{g}$ electrons present around $-$6\,eV.
These states originate from the covalent interaction between
the Mn $e_{g}$ electrons and O 2$p$ states which produce the $e_{g}$ bonding 
(around $-$6\,eV) and antibonding (above $-$2\,eV ) hybrids.
The $t_{2g}$ states are energetically degenerate with the O 2$p$ states in VB
indicating that there are finite covalent interactions between
these states. Owing to this covalent interaction along with the exchange interaction
between the $t_{2g}$ and $e_{g}$ states, a finite DOS of $t_{2g}$ electrons
in the energy range $-$7 to $-$5\,eV is created.
With $E_{F}$ positioned in the middle of the $e_{g}$ band, a JT instability is produced
which causes the oxygen
octahedra to distort and removes the orbital degeneracy.
For the undistorted cubic perovskite structure, the total DOS is shown in Fig.\,\ref{fig:ctdos}
for the nonmagnetic (paramagnetic=P ), F and the A-, C- and G- type AF 
spin configurations. 
Without the JT-caused lattice distortion, the LSDA calculations show that LaMnO$_3$ is metallic
for all these magnetic states. In all
the cooperative magnetic states, the Hund splitting is large and induces an empty minority-spin band
at the Mn site.
The DOS for the P phase given in Fig.\,\ref{fig:ctdos} shows a large value at $E_{F}$.
This is a favorable condition for providing a magnetic splitting and hence the cooperative 
magnetic phase is more stable than the
P phase.
From DOS for the F state of the undistorted structure (see Fig.\,\ref{fig:ctdos}) one
can see that there is a finite number of states present in both spin channels. On the other hand the
half-metallic character clearly appeared in the F state of the orthorhombic structure
(Fig.\,\ref{fig:tdos}). The different hybridization nature of the cubic
and the orthorhombic phases and the noninvolvement of JT effects in the cubic phase
are the possible reasons for the absence of the half-metallic character in the cubic phase.
The total DOS of the AF phases are much different from that of the F phase 
(Fig.\,\ref{fig:ctdos}). Owing to large exchange splitting there is a smaller DOS at the
Fermi level [$N(E_{F})$] for the F and AF phases compared with the P phase.
The calculated value of $N(E_{F})$ for the P phase is 146.6\,states/(Ry. f.u.)
and 16.7, 20.5, 25.1 and 17.8\,states/(Ry f.u.) for A-, C- and G-AF as well as F phases,
respectively.

\subsubsection{Electronic structure of the orthorhombic phase}
Let us now focus on the role of the structural distortion on the
electronic structure of LaMnO$_3$.
It is believed that LaMnO$_3$ is a charge-transfer-type (CT) 
insulator\cite{saitoh95,arima93} according to the Zaanen-Sawatzky-Allen scheme,\cite{zsa85} in which
the lowest-lying gap transition corresponds to the CT excitation from the O 2$p$ to 
Mn 3$d$ state and has four $d$ electrons per Mn$^{3+}$ site with a configuration of 
$t_{2g}^{3}e_{g}^{1}$.  This electronic configuration with one electron
for the two $e_{g}$ orbitals implies that LaMnO$_3$ is a typical JT system.
So, this electronic configuration is susceptible to a strong electron-phonon coupling of
the JT type that splits the $e_{g}$ states into filled $d_{z^{2}}$ and empty
$d_{x^{2}-y^{2}}$, and thus, produces large asymmetric oxygen displacements.
The important factors governing the formation of the electronic structure of 
orthorhombic LaMnO$_3$
are the exchange splitting owing to spin polarization, the ligand field splitting
of $e_g$ and $t_{2g}$ states and the further splitting of the $e_g$ states owing to the JT distortion.
Early theoretical work focused on the undistorted perovskite aristotype structure and it was found that
usual LSDA can not produce the correct insulating ground state for LaMnO$_3$.\cite{pari95}
Using the LAPW method, Pickett and Singh\cite{pickett96} obtained a gap of 0.12\,eV for 
the distorted
LaMnO$_3$ when they include the JT effect and A-AF ordering.
\par
As noticed in earlier studies,\cite{pickett96}
two unexpected behaviors appear when one includes the structural distortion; the 
stabilization of the A-AF over the F phase and an opening of a gap and large
rearrangement of bands.
The valence band DOS of LaMnO$_3$ are derived primarily from Mn 3$d$ and O 2$p$ admixture with dominant
Mn 3$d$ character (Fig.\,\ref{fig:pdos}). 
From the projected density of states (PDOS) it can be seen that the Mn 3$d$ $-$ O 2$p$ hybridization 
is spin dependent. For the case of the
majority-spin channel both Mn 3$d$ and O 2$p$ strongly mix with each other in the whole VB. 
On the contrary, owing to the presence of Mn 3$d$ in the high spin state, the
minority-spin VB is nearly empty for the Mn 3$d$ states. Hence, there is little overlap between
Mn 3$d$ and O 2$p$ states in the minority-spin channel.
Because of CT from La and Mn to the O 2$p$ states, the latter
are almost filled and their contribution to the unoccupied state is minimal. The O 2$s$
states are well localized and are present around $-$18\,eV. 
\par
VB of LaMnO$_3$ is generally composed of four regions. The lowest energy region
contains mainly O 2$s$ bands, above this the La 5$p$ bands are distributed in the
region between $-$17\,eV
and $-$14\,eV. Both the O 2$s$ and La 5$p$ bands are well separated from the bands in the vicinity
of $E_{F}$, and consequently they hardly contribute to the chemical bonding or to transport properties.
The O 2$p$ bands are present in the energy range $-$7\,eV to $-$2\,eV and 
energetically degenerate with the Mn 3$d$ states in the whole VB region indicating 
covalent interaction between Mn and O in LaMnO$_3$. 
The present observation of strong covalency in the ground state of LaMnO$_3$ is 
consistent with the conclusion drawn from photoemission and x-ray-absorption
spectroscopy.\cite{saitoh95} The negligible contribution
of La electrons in the VB region indicates that there is an ionic interaction between
La and the MnO$_6$ octahedra. 
It has been pointed out by Goodenough\cite{goodenough63} that the covalency between
the A site and oxygen is important for the GdFeO$_3$-type distortion.
However our PDOS profile shows that there is only a negligible amount of electrons present 
in VB from the A (La) site indicating that the covalent interaction between
La and O is rather unimportant in LaMnO$_3$. From
PDOS along with the orbital-projected DOS we see that 
both Mn $t_{2g}$ and $e_{g}$ electrons participate in the covalent interaction with
the neighboring oxygens.
The top of VB is
dominated by the majority-spin Mn 3$d$ states indicating the importance of Mn 3$d$ states in 
transport properties such as CMR observed in hole-doped LaMnO$_3$. The bottom of CB
is dominated by the minority-spin Mn 3$d$ electrons and above the La 4$f$ electrons are 
present in a very narrow energy range between 2.6 and 3.5\,eV (Fig.\,\ref{fig:pdos}).
O 3$d$ and Mn 4$p$ states are found around 10\,eV above $E_{F}$. 
As the Mn $d$ states are playing an important role for the magnetism and other physical
properties of LaMnO$_3$, it is worthwhile to investigate these in more detail. 
\par
An approximately cubic crystal field stemming from the oxygen octahedron around Mn 
would split the Mn 3$d$ levels into $t_{2g}$ and $e_{g}$ 
levels. For Mn$^{3+}$ in LaMnO$_3$, three electrons would occupy localized $t_{2g}$
levels, and one electron a linear combination of two $e_{g}$ levels.
The general view of the electronic structure of LaMnO$_3$ is that both $t_{2g}$ 
and e$_g$ orbitals hybridize with O 2$p$ orbitals, 
$t_{2g}$ mainly with 2$p$ $\pi$ and $e_{g}$
mainly with 2$p$ $\sigma$.
The $t_{2g}$ electrons hybridize less with O 2$p$ states and hence may be viewed as local
spins ($S$ = 3/2). In contrast to that, $e_g$ orbitals,
which hybridize more strongly produce rather broad bands. The
strong exchange interaction with the $t_{2g}^{\uparrow}$ subbands along with the JT distortion
lead to the splitting of $e_{g}$ into half occupied $e_{g}^{\uparrow}$ and 
unoccupied $e_{g}^{\downarrow}$ bands.
Our calculations show that $t_{2g}$ bands forms intense peaks in both VB and CB, 
and majority-spin $t_{2g}$
bands are almost completely filled owing to the high spin state of Mn in LaMnO$_3$. 
Contrary to the general opinion, our orbital projected DOS (Fig.\,\ref{fig:ldos}) 
show that both $e_{g}$ 
and $t_{2g}$ electrons are present throughout VB and hence both
types of electrons participate in the covalent bonding with oxygen.
The orbital projected DOS (Fig.\,\ref{fig:ldos}) clearly shows that there is
almost equal amounts of both $e_{g}$ and $t_{2g}$ electrons at the top of VB
and certainly not\cite{jung97} dominated by         
Mn $e_{g}$ bands alone. The DOS in Fig.\,\ref{fig:ldos} $d_{z^{2}}$ states are
shifted to a lower energy than $d_{x^{2}-y^{2}}$ and is more populated.
\par
A comparison of the orbital projected DOS for the Mn $d$ states of the F cubic phase
with that of the AF orthorhombic phase show significant differences.
In the F cubic case the $e_{g}$ levels are well separated from the
$t_{2g}$ levels and there is no $t_{2g}^{\uparrow}$ electrons present in the vicinity
of $E_{F}$. In the A-AF orthorhombic phase both $t_{2g}^{\uparrow}$
and $e_{g}^{\uparrow}$ electrons are present in the vicinity of $E_{F}$. Both $e_{g}$ and
$t_{2g}$ electrons are energetically degenerate throughout the VB range and
in particular the electrons in the $d_{z^{2}}$ orbitals are well localized and degenerate
with the electrons in $d_{yz}$ and $d_{xz}$. Importantly, the JT distortion split the
$e_{g}$ states in the A-AF orthorhombic phase and hence we find semiconducting
behavior.
Photoemission studies\cite{saitoh95} show that the character of the bandgap of LaMnO$_3$
is of the $p$-to-$d$ CT type. Our site projected DOSs predict that
there is considerable $p$ and $d$ character present at the top of VB as well
as in the low energy region of CB. The VB photoemission 
study\cite{saitoh95} has been interpreted to suggest that DOS closer to $E_{F}$ in  
VB contains contributions from Mn $e_{g}^{\uparrow}$. In contrast, our calculations
(Fig.\,\ref{fig:ldos}) show that DOS closer to $E_{F}$ in
VB contains contribution from both Mn $e_{g}^{\uparrow}$ and $t_{2g}^{\uparrow}$
electrons. 
\par
Owing to the JT distortion the $e_{g}$ electrons are subjected to an OO effect in LaMnO$_3$.
As the JT distortion influences the stability of the orbital configuration, one can expect that it 
also should affect the electronic structure of LaMnO$_3$ differently for different 
magnetic orderings. Hence,
we next focus our attention to the electronic structure of orthorhombic LaMnO$_3$ in 
different magnetic configurations. The calculated total DOS for orthorhombic LaMnO$_3$
in the P, F and A-, C- and G-AF arrangements are shown in Fig.\,\ref{fig:tdos}.
From DOS at $E_{F}$ in the P phase, one can deduce pertinent information
concerning the magnetism in LaMnO$_3$. This shows that LaMnO$_3$ is a very favorable compound 
for cooperative magnetism
since it has a large DOS value at $E_{F}$ in the P state 
(Fig.\,\ref{fig:tdos}).
Owing to the splitting of $e_{g}$ states by the JT distortion, a gap opens up 
$E_{F}$ in DOSs
for both the A-AF and G-AF phases (Fig.\,\ref{fig:tdos}).
It is interesting to note that the cubic phase of LaMnO$_3$ is always metallic 
irrespective of the magnetic ordering considered in the calculations. 
Thus the structural distortion is the key ingredient to account for the stabilization of the
insulating
behavior of LaMnO$_3$, consistent with the earlier 
studies.\cite{pickett96} From Fig.\,\ref{fig:tdos} it is interesting to note that 
the P, F and C-AF phases of LaMnO$_3$ are found to exhibit metallic 
conduction even 
when we include the structural distortions in the calculation. This shows that apart
from structural distortion, the AF ordering also plays and important role for stabilizing
the insulating behavior of LaMnO$_3$.
Usually, the $e_{g}$ splitting caused by the JT effect is somewhat underestimated in the ASA 
calculations\cite{solovyev96,pari95,yang00,satpathy96}
and a discrepancy could reflect an uncertainty introduced by the ASA approach.
The energy separation between the $e_{g}$ and $t_{2g}$ levels, caused by the crystal-field 
splitting, is known to be larger than 1\,eV.\cite{okimoto95} Owing to strong overlap
between the $e_{g}$ and $t_{2g}$ levels in the A-AF orthorhombic phase, we are
unable to estimate the crystal field splitting energy. However, owing to the well separated
$e_{g}$ and $t_{2g}$ levels in the cubic phase we estimated this
energy to be $\sim$0.96\,eV. 
The JT distortion lifts the degeneracy of the $e_{g}$ level.
The $e_{g}$ level splitting in our calculation is found to be 0.278\,eV and this is nothing
but the (semiconducting) bandgap in this material.
The Mn $d$ exchange splitting obtained from our calculation is 3.34\,eV and this is found
to be in agreement with 3.5\,eV found by Pickett and Singh\cite{pickett96}
by a LAPW calculation and 3.48\,eV reported by Mahadevan {\em et al.}\cite{mahadevan97}
from LMTO-ASA calculations.
\par
The LMTO-ASA\cite{hu00} calculations reveals that the splittings between the
spin-up and spin-down states on introduction of the orthorhombic distortion
are slightly asymmetric for the two types of Mn.
However, our more accurate full potential calculations
do not show any asymmetry of the splitting and the magnetic moments are
completely canceled due to the AF interaction between the Mn ions. The total DOS in 
Fig.\,\ref{fig:tdos} show clearly that there is a gap opening near $E_{F}$ in the G-AF 
phase with a value of 0.28\,eV, comparable with that of the A-AF phase. As there is
no bandgap in the C-AF and F phases, the above results indicate that the AF coupling
between the layers plays an important role in opening up the bandgap in LaMnO$_3$.
The DOS for the F phase shows a half-metallic feature, viz., the finite total DOS around 
$E_{F}$ comes from one of the spin channels while there is a gap across $E_{F}$ for the other
spin channel. Both spin-up and spin-down e$_g$ states extend over
a wide energy range. Owing to the F ordering within the $ab$ plane in the A-AF phase and
within $ac$ in the C-AF phase, the DOS for these phases bears resemblance to that of the 
F phase. On the
other hand, DOS for the G-AF phase (dominated by the AF superexchange 
interactions) is
quite different from that of the A-, C-AF and the F phases (Fig.\,\ref{fig:tdos}).
Moreover, in the G-AF phase the width of the $e_{g}$ state is narrower than that in the A-, C-AF 
and F phases
and the partially filled $e_{g}$ states in these phases are well separated from the empty $t_{2g}$
states.

\subsection{Magnetic properties}
\label{sec:magnetic}
As the JT coupling between the $e_{g}$ electrons and the distortion modes for the
MnO$_6$ octahedra plays an important role for the physical properties of LaMnO$_3$,
a magnetic-property study of the undistorted cubic phase of LaMnO$_3$ is
important. We have calculated the total energies and magnetic moments for LaMnO$_3$
in the undistorted cubic perovskite structure as well as for the orthorhombic structure
with different magnetic configurations.
The calculated total energies for the P, F, A-, C- and G-AF states of LaMnO$_3$ with the
cubic perovskite structure relative to the A-AF phase with the orthorhombic structure is given in
Table\,\ref{table:de}. According to these data the cubic phase of LaMnO$_3$ should be stabilized
in the F phase.
Stabilization of the F phase in the cubic structure concurs
with earlier findings.\cite{hu00,pickett96}
The LAPW calculations of Hamada {\em et al.}\cite{hamada95} for undistorted LaMnO$_3$  
show that the A-AF phase is 1\,eV above the F phase, whereas Pickett
and Singh\cite{pickett96} found a difference of only 110\,meV. Our calculations shows that the A-AF
phase is 60\,meV above the F phase in the cubic perovskite structure (Table\,\ref{table:de}).
We used GGA, SO coupling and a large number of {\bf k} points along with a
well converged basis set. This may account for the difference between the present work and
the earlier studies. Our orbital projected DOS for the $d$ electrons of Mn (Fig.\,\ref{fig:cldos})
show that the $e_{g}$ electrons are only distributed in the vicinity of $E_{F}$
in the majority-spin channel. 
The removal of the JT distortion enhances the exchange interaction originating from the $e_{g}$ states
drastically and also reduces the negative exchange from the $t_{2g}$ state. Hence, the total
interplane exchange interaction is positive and the system stabilizes in the F phase.
\par
The calculated total energies of various magnetic configurations for LaMnO$_3$ in the
orthorhombic structure (Table\,\ref{table:de}) shows that 
the orthorhombic structure with the A-AF ordering of the moments is the ground
state for LaMnO$_3$. 
The stabilization of the A-AF state is 
consistent with the neutron diffraction findings.\cite{wollan55,koehler57}
Further, our calculations predict that the F phase 
is only 24\,meV higher in energy than the ground state. On the other hand, the P
phase is $\sim$\,1.4\,eV higher in energy than the ground state. 
From the total energy of the various possible magnetic arrangements it is clear that intralayer
exchange interactions in LaMnO$_3$ are F and considerably stronger than
the AF interlayer couplings, the latter being very sensitive to lattice distortions.
This observation is in agreement with the conclusion of
Terakura {\em et al.}\cite{terakura99} from electronic structure studies and in
qualitative agreement with inelastic neutron scattering results.\cite{hirota96,moussa96}
The present observation of a small energy difference between the A-AF and F phases
is consistent with earlier theoretical studies.\cite{hotta99}
It should be noted that LMTO-ASA calculations\cite{hu00} predict the F state to be
lower in energy than the AF state also for the orthorhombic structure (Table\,\ref{table:de}).
However, the LSDA+$U$ calculations\cite{hu00} with lattice distortion yielded the correct ground state.
Hence, it has been concluded that the correlation effect plays an important role for obtaining
the correct electronic structure of LaMnO$_3$. 
It should also be noted that the theoretically optimized crystal-structure theory
predicts\cite{sawada97} that F is more stable than the observed A-AF state in
LaMnO$_3$. On the other hand, our calculations, without
the inclusion of the correlation effect, give the correct ground state for LaMnO$_3$ indicating
that the correlation effect is not so important in this case.
\par
If the interlayer F coupling is stronger than the intralayer AF coupling the system
will be stabilized in the C-AF phase. However, our calculations show that C-AF phase is
higher in energy than the A-AF and F phases.
Due to the JT instability of Mn $d^{4}$, a substantial
energy gain is expected when we include the actual structural distortions in our calculations.
Hence, a gain of 0.323\,eV/f.u. is found for F and 
0.407\,eV/f.u. for A-AF (Table\,\ref{table:de}) when the structural
distortions were included. 
The magnetic ordering is such that the difference in total energy between F 
and A-AF (C-AF) give information about the exchange-coupling energy
within the plane (perpendicular to the plane for C-AF). Our calculations show that
the AF intraplane exchange interaction energy is smaller (24\,meV for the orthorhombic
phase and 59\,meV for the cubic phase) than the AF interplane exchange interaction
energy (41\,meV for the orthorhombic phase and 92\,meV for the cubic phase). The present
findings are consistent with the experimental studies in the sense that 
neutron scattering measurements\cite{hirota96} on a LaMnO$_3$ single crystal 
showed a strong intraplane F coupling and a weak AF interplane
coupling.
\par
Now we will try to understand the microscopic origin for the stabilization of A-AF in LaMnO$_3$.
In the ideal cubic lattice the hybridization between the Mn $t_{2g}$ and $e_{g}$ states
is nearly vanishing. As expected for a half-filled $t_{2g}$ band, the $t_{2g}$-type
interatomic exchange facilitates AF ordering.
Our orbital-projected DOS show that the $d_{x^{2}-y^{2}}$ electrons are distributed
in the whole VB region and are mainly populated in the top of VB.
The JT distortion induces orbital ordering in which the orbitals confined to the $ab$ plane 
($d_{x^{2}-r^{2}}$ and $d_{y^{2}-r^{2}}$) are
dominantly populated and the counter $e_{g}$ orbitals of $d_{z^{2}}$ symmetry are less
populated in the vicinity of $E_{F}$. In this case itinerant band ferromagnetism is operational
in the $ab$ plane and is responsible for the F ordering with in the plane. The electrons
in the $d_{yz}$ and $d_{z^{2}}$ orbitals are well localized and these make the interplane exchange
interaction AF, which in turn stabilizes the A-AF phase. However, the
electrons in the $d_{xz}$ orbitals are well delocalized and have almost the same
energy distribution as $d_{x^{2}-y^{2}}$. This weakens the AF coupling between the layers 
and in turn makes the energy difference between the F and A-AF phases
very small. 
The CMR effect in manganites may be understood as follows. When charges are localized by 
strong electron-electron interactions the system becomes an AF insulator,
but this state is energetically very close to the metallic F state in 
LaMnO$_3$. Consequently, aligning of spins with an external magnetic field
will activate the metallic F states and cause a large gain of kinetic
energy, i.e. a CMR phenomenon.
\par
Table\,\ref{table:moment} lists the calculated magnetic moment at the Mn site in LaMnO$_3$
for different spin configurations in the undistorted cubic perovskite structure as well as the
distorted orthorhombic structure, including for comparison corresponding values from 
other theoretical studies and experimental neutron diffraction
results.
Without hybridization, the Mn spin moment should take an appropriate integer value 
(4\,$\mu_{B}$/Mn atom in the
case of high spin state) and the oxygen moment should be negligible. Owing to the covalent
interaction between the Mn $d$ and O $p$ states, experimental as well as 
theoretical studies will give smaller Mn moments than predicted by the ionic model. 
For the F phase in the orthorhombic structure, Mn polarizes the 
neighboring oxygens and the 
induced moment at the O(1) site is 0.07\,$\mu_{B}$/atom and 
at the O(2) 0.059\,$\mu_{B}$/atom and these moments are coupled ferromagnetically 
with the local 
moments of Mn$^{3+}$. There are small differences in the magnetic moments
for the different magnetic arrangements indicating that these moments have a distinct
atomic-like character. From Table\,\ref{table:moment}  it should be noted that the 
LMTO-ASA approach generally gives
larger moments than the accurate full-potential calculation. The reason for
this difference is that the full-potential calculations estimate the moments using the spin density
within the muffin-tin spheres so that the spin-density in the interstitial region is
neglected. Our calculated magnetic moment at the Mn site
in the A-AF orthorhombic structure is comparable with the experimental
value (Table\,\ref{table:moment}).
\par
The electric field gradient for Mn in the F phase obtained 
by the FLAPW calculation is 3.579 $\times$ 10$^{21}$ V/m$^{2}$ as compared with 
$-$1.587 $\times$ 10$^{21}$ V/m$^{2}$ for the A-AF phase. The asymmetry parameter $\eta$ of the
electric field gradient tensor follows from the relation 
$\eta = \frac{V_{xx} - V_{yy}}{V_{zz}}$, which gives $\eta$=0.95
for Mn in the A-AF phase  
in good agreement with $\eta$ = 0.82 obtained from low temperature
perturbed-angular-correlation spectroscopy.\cite{rasera98}
The magnetic hyperfine field at the Mn site derived from our FLAPW calculation
is 198\,kG for the A-AF phase and 176\,kG for the F phase.
\par
A comment on the appearence of the F state by
doping of divalent elements in LaMnO$_3$ is appropriate. 
It has long been believed that the appearence of the F ground sate in 
metallic La$_{1-x}${\it AE}$_x$MnO$_3$ can be explained by the double-exchange interaction of Mn$^{3+}$
($t_{2g}^{3} e_{g}^{1}$) and Mn$^{4+}$ ($t_{2g}^{3} e_{g}^{0}$).
Recent experiments and theoretical investigations have revealed many discrepancies in the 
simple double-exchange model. Urushbara {\em et al.}\cite{urushibara95} pointed out that
the simple double-exchange model is methodologically inappropriate for predicting the AF
insulating phase (0.1 $\leq x \leq$ 0.15, $T$ $<$ $T_C$),
for the canted AF insulating phase with large canting angle
or large magnetic moment estimated by Kawano {\em et al.}\cite{kawano96} in the F phase region. 
Millis {\em et al.}\cite{millis95} proposed
that in addition to the double-exchange mechanism, a strong electron-phonon interaction arising
from the JT splitting of the outer Mn $d$ level is important.
As our calculations predict the cubic phase of LaMnO$_3$ to be the F state,
there must be another mechanism than the doping by the divalent elements which reduces the
JT distortion and in turn increases the interlayer exchange coupling to
stabilize the F state. Moreover, our calculations show that within
the $ab$ plane the Mn ions are F coupled 
without involvement of Mn in different ionic states. If the double-exchange 
mechanism is operational in LaMnO$_3$
one can expect a charge imbalance between the Mn atoms in the $ab$ plane (with F alignment) 
This suggests that the F coupling in LaMnO$_3$ can
be explained within the frame work of itinerant band ferromagnetism.

\subsection{Magneto-optical properties}
\label{sec:moke}
The magneto-optical (MO) Kerr-effect can be described by the off-diagonal elements of the
dielectric tensor which in a given frequency region originates from optical
transitions with different frequency dependence for right and left circularly
polarized light caused by the SO splitting of the states involved.\cite{shen64}
Considerable interest has recently been focused on studying MO properties of
materials owing to their potential for application in rewritable high-density data storage. 
For magneto-optical information storage one requires materials with a large Kerr-effect
as well as perpendicular magnetic anisotropy. Uniaxial magnetic anisotropy is observed
for La$_{0.67}$Ca$_{0.33}$MnO$_3$ films grown on SrTiO$_3$ [001] substrates\cite{donnell98} 
and confirmed theoretically.\cite{shick99}
\par
The energy difference between the A-AF and F states of LaMnO$_3$ is very
small and hence a small magnetic field could drive a transition from A-AF to F.
Also, it is interesting to note that LaMnO$_3$ produced by 
annealing in a oxygen
atomosphere shows\cite{huang97} a simple F structure below $T_{C}$ = 140\,K, with moments oriented
along $c$.
Manganites exhibiting the CMR effect are also in the F state and hence it is interesting to study the
F phase of LaMnO$_3$ in more detail. 
Polar Kerr-rotation measurements on hole doped LaMnO$_3$ show that Bi doping
enhances the Kerr-rotation to a value of 2.3\,$^o$ at wavelength of 0.29\,nm at 78\,K.\cite{popma75}
It may be recalled that the large MO effect in MnPtSb is usually linked with its half-metallic
behavior.\cite{half} Hence, as the F phase of LaMnO$_3$ also
possesses half-metallic behavior with a bandgap of 2.38\,eV (between top of the
minority-spin VB and the bottom of CB in Fig.\,\ref{fig:tdos} at $E_{F}$), it is interesting to examine its MO properties.
\par
In the case of F state LaMnO$_3$, our calculations predicted half-metallic behaviour and hence
the intraband contribution will be of importance to predict the MO property reliably. 
Therefore, we have taken into account the intraband contribution
using the Drude formula with the same relaxation time as earlier.\cite{ravi99} 
We have calculated the unscreened plasma frequency by integrating over
the Fermi surface using the FLAPW method.
For F state La$_{0.7}$Ca$_{0.3}$MnO$_3$ Pickett and Singh predicted\cite{pickett97} a 
plasma frequency of 1.9\,eV. 
Several groups have reported an anomalously small Drude weight in both 
La$_{0.7}$Ca$_{0.3}$MnO$_3$ and La$_{0.7}$Sr$_{0.3}$MnO$_3$.\cite{kim98,okimoto95}
Interpreting the latter findings in terms of enhanced optical mass, the 
effective mass values established optically are much greater than those
derived from the specific-heat measurements.\cite{hamilton96} The theoretically
established half-metallic feature of the F state in LaMnO$_3$ suggests that the small Drude
weight is originating from large exchange splitting which contracts the Drude contribution
coming from the minority-spin channel.
In accordance with the half-metallic nature of the F state, our 
calculated DOS at $E_{F}$ is very small [3.36\,states/(Ry f.u.)]
As a result, the calculated unscreened plasma frequency along 
$a$, $b$ and $c$ are 1.079, 1.276 and 0.926\,eV, respectively. 
Owing to the half-metallic nature of this material there is no intraband contribution arising from
the majority-spin channel. Hence, we conclude that the experimentally observed\cite{kim98,okimoto95}
anomalously small
Drude weight in La$_{0.7}$Ca$_{0.3}$MnO$_3$ and La$_{0.7}$Sr$_{0.3}$MnO$_3$
is due to half-metallic behavior. Our calculated MO spectra for F state LaMnO$_3$ is 
given in Fig.\,\ref{fig:moke}. Our recent MOKE studies\cite{ravi} of FePt shows that Kerr-rotation 
spectra can be reliably predicted even up to as high energies as 10\,eV with
the formalism adopted here. Hence, we have depicted the calculated MO spectra up to
8\,eV in Fig.\,\ref{fig:moke}. 
The Kerr-rotation ($\Theta_{K}$) shows a positive or negative peak when the Kerr-ellipticity 
($\eta_{K}$) passes through zero.
The frequency dependent Kerr-rotation and ellipticity spectra
are experimentally established\cite{yamaguchi98,popma75} for hole-doped LaMnO$_3$. As
there are no experimental MO spectra available for comparison with our theoretical spectra
for (pure) F state LaMnO$_3$,
we have made comparison 
with the experimetal spectrum for hole-doped LaMnO$_3$ (Fig.\,\ref{fig:moke}). 
\par
MO effects are proportional to the product of the SO-coupling strength
and the net electron-spin polarization. This makes MO effects sensitive
to the magnetic electrons, i.e. the 3$d$ electrons of Mn in LaMnO$_3$. 
In order to promote the understanding of the microscopic origin of the MO 
effect, we present the
off-diagonal elements of the imaginary part of the dielectric tensor in Fig.\,\ref{fig:e2}. Note that
the $\epsilon_{2}^{xy}$ spectra can have either positive or negative sign since it is proportional
to the difference in the absorption of RCP and LCP light. The sign of $\epsilon_{2}^{xy}$
is thus directly related to the spin polarization of the electronic states that contributes
to the MO effect.
Owing to the half-metallic behavior of the F state of LaMnO$_3$ the off-diagonal components
of the dielectric tensor below 2.46\,eV originates only from the majority-spin electrons
(see Fig.\,\ref{fig:e2}).
So, the peaks at 0.4, 0.9 and 2\,eV in the Kerr-rotation spectra stem from
transition from Mn $e_{g}^{\uparrow}t_{2g}^{\uparrow}$ to the hybridized Mn$-$O   
Mn($e_{g}t_{2g}-$O 2$p$)$^{\uparrow}$ bands.
The experimental $\epsilon_{2}^{xy}$ spectra show\cite{yamaguchi98}
two peaks, one at
1.2\,eV and the other at 3.5\,eV. These two peaks are assigned to be $t_{2g}^{3}e_{g}^{1}
\rightarrow t_{2g}^{3}e_{g}^{2}L$ transitions involving $t_{2g}^{4}e_{g}^{1}L$ minority-spin electrons.
electrons, respectively. The absence of the 3.5\,eV 
peak in our theoretical spectra indicate that it is an effect of the hole-doping.
\par
It is experimentally observed that Kerr-rotation spectra
change significantly with the temperature\cite{popma75} as well as with the doping 
level.\cite{yamaguchi98} Our calculated MO spectra are valid only for the stoichiometric
F state LaMnO$_3$ at low temperature. 
The polar MO Kerr-rotation data for La$_{0.7}$Sr$_{0.3}$MnO$_3$ at 78\,K
was taken from the experimental spectra of Popma and Kamminga\cite{popma75}.
Even though the experimental MO spectra La$_{0.8}$Sr$_{0.2}$MnO$_3$ were 
measured\cite{yamaguchi98} at 300\,K with a magnetic field of 2.2\,kOe, our calculated
Kerr-rotation spectrum is comparable in the lower
energy region (Fig.\,\ref{fig:moke}). The discrepancy between the experimental
and theoretical Kerr-spectra in the higher-energy region may be explained as a temperature and/or
hole doping effect. We hope that our theoretical findings may motivate to
measurements on fully magnetized LaMnO$_3$ at low temperatures.
Solovyev {\em et al.}\cite{solovyev97} have performed theoretical calculation on MO 
properties such as Kerr-rotation and ellipticity for LaMnO$_3$ for different 
canted-spin configurations. 
As the F state is only 24\,meV above the ground state, it is quite
possible that the F state can be stabilized experimentally. However, since we have 
studied the MO properties
of F state LaMnO$_3$ our result can not be compared with the findings of 
Solovyev {\em et al.}\cite{solovyev97}
\par
Lawler {\em et al.}\cite{lawler94} found strong Faraday rotation at 
1.5\,eV ($\theta_{F} >$ 1 $\times$ 10$^{4}$ $^{o}$/cm) and 3\,eV ($\theta_{F}
>$ 4.10$^{4}$ $^{o}$/cm) for La$_{1-x}$Ca$_x$MnO$_3$
when (0.2 $\leq x \leq$ 0.5). Our theoretical Faraday rotation spectra also show two
prominent peaks at 1.5 and 4\,eV, the latter having the highest value ($\sim$ 
1.25$\times$10$^{5}$ $^{o}$/cm) in the spectrum. The lower-energy peak in the
experimental spectrum\cite{lawler94} of La$_{1-x}$Ca$_x$MnO$_3$ is interpreted as
associated with both ligand-to-metal charge transfer and $d$-$d$ transitions. However, there are
as yet no experimental frequency-dependent Faraday rotation and ellipticity measurements 
available.

\subsection{Optical properties}
\label{sec:optic}
A deeper understanding of optical properties is important from a fundamental point of view, since
the optical characteristics involves not only the occupied and unoccupied part of the 
electronic structure but also the character of the bands.  So, in order to compare 
the band structure
of LaMnO$_3$ directly with experimental facts\cite{arima93,jung97,takenaka}
we have calculated the optical spectra for LaMnO$_3$ (for earlier theoretical studies
see Refs.\,\onlinecite{solovyev96,bouarab96}).
The experimental reflectivity spectra are rather confusing and
controversial (see the discussion in Ref. \onlinecite{takenaka}). Recent 
experiments\cite{takenaka} indicate
that the optical spectrum of the manganites is very sensitive to the condition at
the surface.
As the optical properties of materials originate from interband transitions
from occupied to unoccupied bands it is more instructive to turn to the
electronic energy-band structure. The calculated energy-band structure of LaMnO$_3$ 
in the A-AF state orthorhombic structure is shown in Fig.\,\ref{fig:bnd}. From this
illustration it is immediately clear that LaMnO$_3$ is an indirect bandgap semiconductor 
where the bandgap is between the S-Y direction on the top of VB and the
$\Gamma$ point at the bottom of CB. There are two energy bands present in
the top of VB, well separated from the rest of the VB.
Our detailed analysis shows that these two energy bands have mainly Mn $d_{x^{2}-y^{2}}$
and $d_{xz}$ character with a small contribution from O $p$. The La 4$f$ electrons contribute with
a cluster of bands between 2.5 to 3.5\,eV in CB (Fig.\,\ref{fig:bnd}). The Mn 3$d$ 
and O 2$p$ electrons are distributed over the whole energy range of VB.
\par
Saitoh {\em et al.}\cite{saitoh95} reported strong covalency and suggested that the energy
gap  in
LaMnO$_3$ should be considered as of the CT type.
The bandgap estimated from our DOS studies for A-AF state LaMnO$_3$ is 0.278\,eV 
and this is found to be in good agreement with the value of 0.24\,eV obtained from
resistivity measurements by Mahendiran {\em et al.}\cite{mahendiran96} whereas
Jonker\cite{jonker54} reported 0.15\,eV. However, our 
value for the direct gap
(0.677\,eV) between occupied and empty states at the same location in BZ 
is too low in comparison with optical
(1.1\,eV)\cite{arima93} and photoemission (1.7\,eV)\cite{saitoh95} measurements.
(It should be noted however that optical gaps are usually defined at the onset of 
an increase in spectral
intensity in the measured optical variable.) It is also useful to compare our calculated
bandgap with other theoretical results. The LSDA and LSDA+$U$ calculations of
Yang {\em et al.}\cite{yang99} using the LMTO-ASA method gave a bandgap of 0.1 and 
1.0\,eV, respectively.
The LSDA+$U$2 approach (where $U$ is applied only to the $t_{2g}$ electrons)\cite{solovyev96}
yielded a bandgap of 0.2\,eV. Hence, our bandgap is somewhat
larger than that of other LSDA calculations. On the other hand, Hartree Fock 
calculations\cite{mizokawa96} gave an unphysically large gap (3\,eV) for LaMnO$_3$.
\par
As all the linear optical properties can be derived from $\epsilon_{2}(\omega)$ 
we have illustrated this quantity in Fig.\,\ref{fig:e2} and we 
have compared our theoretical spectra with the experimental $\epsilon_{2}(\omega)$
spectra derived from reflectivity measurements. 
The illustration shows that our calculated spectra are in good agreement with the 
experimental data
at least up to 20\,eV. This indicates that unlike earlier reported $\epsilon_{2}$ spectra
obtained from ASA calculations\cite{solovyev96,bouarab96}, accurate full-potential calculations
are able to predict the electronic structure of LaMnO$_3$ reliably not only for the
occupied states but also for the excited states. It is interesting to note that we are
able to predict correctly the peaks around 1.5, 4.7, 8.8 and 20\,eV without the 
introduction of so-called 
scissor operations. This suggests that electron-correlation effects are less
significant in LaMnO$_3$.
The peak around 4.7\,eV is reasonably close to the experimental feature reported
by of Arima {\em et al.}\cite{arima93} (Fig.\,\ref{fig:e2}). Our theoretical peak at 8.8\,eV
in the spectra is in quantitative agreement with both sets of experimental data included
in the figure. Our calculations are also able to predict correctly the experimentally observed
peak at 20\,eV by Jung {\em et al.}\cite{jung97} (which Arima
{\em et al.}\cite{arima93} failed to record).
\par
Now we will try to understand the microscopic origin of the optical interband transitions in
LaMnO$_3$. The peak around 1.5\,eV has been assigned\cite{jung97} as intra-atomic 
$e_{g}^{1}(Mn^{3+}) \rightarrow e_{g}^{2}(Mn^{3+})$ transitions.
Note that such $d \rightarrow d$ transitions are not allowed
by the electric dipole selection rule, but it has been suggested\cite{jung98} that a strong
hybridization of the $e_{g}$ bands with the O 2$p$ bands will make such 
$d \rightarrow d$ transitions optically active. 
The interband transition between JT split bands gives low-energy transitions
in the $\epsilon_{2}(\omega)$ spectra and the sharp peak features present in 
$\epsilon_{2}\|a$ and $\epsilon_{2}\|c$ are attributed to such interband transition.
The partial DOS 
for LaMnO$_3$ given in Fig.\,\ref{fig:pdos} show that there is a considerable
amount of O 2$p$ states present at the top of VB as well as at the bottom of CB 
arising from strong covalent interaction between Mn $d$ and O $p$ states. Hence, we
propose that the lower energy peak in the $\epsilon_2(\omega)$ spectra 
originates from 
[(Mn $d_{x^{2}-y^{2}}$ $d_{xz}$;O $p$ hybridized) $\rightarrow$ (Mn $d$;O $p$)]
optical interband transition.
The peak around 4.7\,eV in the $E\|b$ $\epsilon_{2}(\omega)$ spectrum originates mainly from 
hybridized (Mn $d_{z^{2}}$,O $p$) states to unoccupied hybridized (Mn 3$d$;O $p$)  states.
All the majority-spin $d$ electrons participate in the interband transition from
3\,eV up to 10\,eV in in Fig.\,\ref{fig:e2}.
\par
In the cubic case, $\epsilon_{2}(\omega)$ is a scalar, and in the orthorhombic
case it is a tensor. So we have calculated the dielectric component with the light
polarized along $a$, $b$ and $c$ as shown in Fig.\,\ref{fig:e2}. 
The optical anisotropy in this material can be understood from our directional-dependent,
optical dielectric tensor shown in Fig.\ref{fig:e2}. 
From the calculations we see that the $\epsilon_{2}(\omega)$ spectra along  
$a$ and $c$ are almost the same. 
Large anisotropy is present in
the lower energy region of the $\epsilon_{2}(\omega)$ spectra. In particular the sharp
peaks present for $E\|a$ and $E\|c$ which is less pronounced for
$E\|b$.
The optical conductivity obtained by Ahn and 
Millis\cite{ahn00} from tight-binding parameterization of the band structure also
shows a sharp peak feature in the lowest energy part of $\sigma_{xx}$
whereas this feature is absent in their $\sigma_{zz}$ spectrum.
Our calculation predicts that the interband transition due to JT splitting is less
pronounced in the $\epsilon_{2}$ spectrum corresponding to $E\|b$. In order to confirm 
these theoretical predictions polarized optical property measurements on LaMnO$_3$ are needed.
\par
Soloveyev\cite{solovyev} suggested that the optical anisotropy in 
A-AF state LaMnO$_3$ is due to two factors:
(i) Owing to large exchange splitting the minority-spin $d_{z^{2}}$ states near
$E_{F}$ will contribute less to $\epsilon_{2}^{xx}$. (ii) Owing to 
A-AF ordering and JT distortion, the contribution of $d_{z^{2}}$ character
to the states with $d_{x^{2}}$ and $d_{z^{2}}$ symmetry is significantly
reduced and hence the intensity of $\epsilon_{2}^{xx}$ in the low-energy region
should also be reduced.
In order to understand the role of the JT distortion for the optical properties 
of LaMnO$_3$, we  
calculated also the optical dielectric tensor for LaMnO$_3$ in the cubic perovskite structure
with A-AF ordering (Fig.\,\ref{fig:e2}). Just like for the orthorhombic A-AF phase,
there is large optical anisotropy present in the lower-energy ($<$ 2\,eV) part of the 
spectrum. 
This indicates that the  A-AF spin ordering is responsible for the large anisotropy 
in the optical dielectric tensor which in turn is a result of different spin-selection 
rules applicable to in-plane
and out-of-plane optical transitions. 
However, the optical anisotropy
in the orthorhombic A-AF phase is much larger than the corresponding undistorted cubic phase  
(Fig.\,\ref{fig:e2}) indicating that apart from A-AF ordering, the JT distortion also 
contributes to large
optical anisotropy in the lower energy region of the optical spectra of LaMnO$_3$.
\par
In order to compare our calculated spectra with those obtained by the
LMTO-ASA method, we have plotted the optical dielectric tensors obtained by 
Solovyev {\em et al.}\cite{solovyev96} from LSDA calculation in Fig.\,\ref{fig:e2}. 
Our calculated $E\|b$ spectrum below 2\,eV is found to be in good
agreement with the $E\|xx$ and $E\|yy$ spectra of these authors. However, their $E\|zz$
spectra is much smaller in the lower-energy region than our results
as well as the experimental
findings (Fig.\,\ref{fig:e2}).
From the optical spectrum calculated by the LDA+$U$ method, Solovyev 
{\em et al.}\cite{solovyev96} found
that the low-energy part (up to 3\,eV) reflects excitations from
O $p$ to an unoccupied band formed by alternating Mn $d_{x^{2}-z^{2}}$ and $d_{y^{2}-z^{2}}$ 
orbitals. However, their study poorly described the higher-energy excitations. This
discrepancy may be due to the limitations of ASA or the minimal basis they have
used in the calculation (La 4$f$ states where not included).
Bouarab {\em et al.}\cite{bouarab96} calculated optical conductivity spectra 
for LaMnO$_3$. They used the 
LMTO-ASA method where nonspherical contributions to the potential are not
included. Further, these calculations are made with a minimal basis set and for the 
undistorted cubic
perovskite structure rather than the actual distorted perovskite structure. Hence, it
does not make sense to
compare our calculated optical spectra directly with the latter findings.
\par
As the linear optical spectra are directly measurable experimentally, we have reported
our calculated optical spectra in Fig.\,\ref{fig:optic}.
Also, to understand the anisotropy in the optical properties
of LaMnO$_3$ we show the linear optical spectra for LaMnO$_3$ along the $a$ and 
$b$ axis in the same illustration. A large anisotropy in the optical properties for the low-energy
region is clearly visible in the reflectivity as well as the extinction coefficient spectra.
(Fig.\,\ref{fig:optic} shows the experimentally
measured reflectivity by Jung {\em et al.}\cite{jung97}, Arima {\em et al.}\cite{arima93}
and Takenaka {\em et al.}\cite{takenaka99} in comparision with calculated
spectra from our optical dielectric tensors.). The reflectivity
measured by Takenaka {\em et al.} above 8.4\,eV is found at a higher value than 
found in the spectra of other two experiments (Fig.\,\ref{fig:optic}). 
Overall our calculated reflectivity spectra are found to be in good qualitative
agreement with the experimental spectra up to 30\,eV in Fig.\,\ref{fig:optic}. The
peaks around 10 and 25\,eV in our reflectivity spectra concur with the findings of 
Takenaka {\em et al.}.\cite{takenaka99} 
As the optical properties of LaMnO$_3$ are anisotropic,
it is particularly interesting to calculate the effective number of valence electrons,
$N_{eff}(\omega)$, participating in the optical transitions in each direction.
Hence, we have calculated (see Ref.\,\onlinecite{ravi99}) the number of
electrons participating in the optical interband transitions in different crystallographic
directions,
and the comparison between
$N_{eff}(\omega)$ the $E\|a$ and $E\|b$ shows significant differences in the low-energy 
(Fig.\,\ref{fig:optic}).
Except for the reflectivity 
there is no optical properties measurements for LaMnO$_3$ available.
Hopefully our findings will motivate such studies.

\subsection{XPS and BIS studies}
In spite of intense experimental\cite{abbate92,park96,saitoh95,jung97,chainani93}
and theoretical\cite{solovyev96,sarma95,satpathy96,pickett96,sarma95} efforts to understand the
electronic structure of LaMnO$_3$, there are still a number of ambiguities concerning the VB
features. Both x-ray (XPS) and ultraviolet photoemission spectroscopy (UPS)
experiments showed a double peak structure between $-$10\,eV and  
$E_{F}$.\cite{park96,saitoh95,chainani93} Even though it is widely accepted that the
double-peak structure arises from O $p$ and Mn $d$ bands, an important controversy still remains
in that some authors\cite{satpathy96,sarma95,saitoh95} have argued that the O $p$ band lies 
below the Mn $t_{2g}$ bands, while others\cite{jung97} have suggested the opposite.
In Fig.\,\ref{fig:xps} we have compared our calculated value of VB XPS intensity
with the experimental spectra.\cite{saitoh95}
In all the experimental studies, the main VB features have a marked increase in binding energy around
1.5$-$2\,eV, remaining large in the region 2$-$6\,eV, and falling off in the
range 6$-$8\,eV. 
\par
The experimental XPS data show three peak intensity features between
$-$7 and $-$3\,eV. As can be seen from Fig.\,\ref{fig:xps}, these three peaks are well
reproduced in the calculated profile. The large experimental background intensity makes a direct
comparison with the calculated peak feature around $-$1.8\,eV  
difficult.
The overall agreement between the theoretical and experimental positions of peaks and shoulders 
in the XPS spectra is very satisfactory
(Fig.\,\ref{fig:xps}).
Note that the experimental XPS spectra
does not exhibit any appreciable intensity in a correlation-induced satellite at 
higher-binding energies in contrast to intense satellite features usually found in transition
metal monoxides, e.g. NiO.
Thus, the good agreement between experimental and theoretical spectra indicates that the
on-site Coulomb correlation effect is not significant in LaMnO$_3$. 
\par
Knowledge of the theoretically calculated  
photoemission spectra (PES) has the advantage that one can identify contributions
from  different constituents
to the PES intensity in different energy ranges. From Fig.\,\ref{fig:xps} it is
clear that the XPS intensity in the energy range $-$3.5\,eV to $E_{F}$ mainly 
originates from Mn $d$ states. The O $p$ electrons contribute to the PES intensity in 
the energy range
$-$7 to 3.5\,eV. Below $-$4\,eV both La and Mn atoms contribute equally to the PES
intensity. Pickett
and Singh\cite{pickett96} compared their theoretical DOS with PES and found
significant differences. Thus, the good agreement between the experimental and our theoretical XPS
indicates that the matrix-element effect is important in LaMnO$_3$.
\par
We now turn to a discussion of the BIS spectra. The physical process in BIS is as follows.
An electron of energy $E_{1}$ is slowed down to a lower energy level $E_{2}$ corresponding
to an unoccupied valence state of the crystal. The energy difference is emitted as 
bremsstrahlung radiation, which reflects the density of unoccupied valence states.\cite{speier85} 
Our calculated BIS spectrum is compared with the experimental spectrum\cite{sarma96}
in Fig.\,\ref{fig:xps} which show that theory is able to reproduce the experimentally observed
peak at 2\,eV.  There is an intense peak above 3.5\,eV in the experimental
spectrum which originates from the La 4$f$ electrons, but owing to its large intensity 
this is not shown in Fig.\,\ref{fig:xps}.

\subsection{XANES spectra}
Another experimental technique that provides information about CB is
x-ray absorption spectroscopy. 
Calculated O $K_{\alpha}$, Mn $K_{\alpha}$ and Mn $L_{II,III}$ XANES 
spectra for LaMnO$_3$ are shown in Fig.\,\ref{fig:xanes}. Because
of the angular momentum selection rule (dipole approximation), only O 2$p$ states and
Mn 4$p$ states contribute to the O $K$-edge and Mn $K$-edge spectra, respectively. The Mn 4$s$ 
and Mn 3$d$ states contribute to the Mn $L_{II,III}$ spectrum.
The calculated X-ray absorption spectrum for the O $K$ edge (VB $\rightarrow$ 1$s$)  
is given in the upper
panel of Fig.\,\ref{fig:xanes} along with available experimental data.
The experimental spectrum has two prominent peaks in the low-energy
range and both are well reproduced in our theoretical spectra even without including
the core-hole effect in the calculation. As reflected in different Mn-O distances
there are two kinds  (crystallographically) of oxygens present in
LaMnO$_3$ and results in considerable differences
in peak positions in the XANES spectra. 
\par
The calculated $L_{II,III}$ spectrum
is shown in the middle panel of Fig.\,\ref{fig:xanes}. Due to the nonavailability
of experimental Mn $L_{II,III}$ spectra for LaMnO$_3$ we have compared our calculated 
spectrum with
that experimentally established\cite{kurata93} for Mn$^{3+}$ in Mn$_{2}$O$_3$.
The calculated Mn $K$-edge spectrum for LaMnO$_3$ is given in the lower panel of 
Fig.\,\ref{fig:xanes} along with the available experimental $K$-edge spectra. 
Although DOS are large for the empty $e_{g}^{\uparrow}$ and 
$t_{2g}^{\downarrow}$ bands, they are not directly accessible in $K$-edge absorption
via dipole transitions.
Thus the main $K$-edge absorption begins where the large part 
of the $4p$ states are highly delocalized and extend over several Mn atoms.
Takahashi {\em et al.}\cite{takahashi99} show that the intensity of the Mn $K$-edge 
spectrum increases
with an increasing degree of local lattice distortion and is insensitive to the
magnetic order.
The experimental Mn $K$-edge spectra show two peaks in the lower energy region and
these are well reproduced in our theoretical spectrum. In particular our
Mn $K$-edge spectrum is found to be in very good agreement with that of 
Subias {\em et al.}\cite{subias98} in the whole energy range we have considered.
The good agreement between experimental and calculated XANES spectra 
further emphasizes the relatively little significance of correlation effects in LaMnO$_3$ and also
shows that the gradient-corrected full-potential approach is able to predict
even the unoccupied spectra quite correctly.

\section{SUMMARY}
\label{sec:con}
Transition metal oxides are generally regarded as a strongly correlated systems which 
are believed to be not properly treated by electronic structure theory. The present 
paper has demonstrated, however,
that accurate full-potential calculations including SO coupling and  
generalized-gradient corrections can account well for several features 
observed for LaMnO$_3$.
We have presented a variety of results based on the {\em ab initio} local spin-density approach
that are broadly speaking in agreement with experimental results for both ground and 
excited state properties.
The present study strongly support 
experimental\cite{zhao99,billinge96,jaime96,palstra97,booth98} 
and theoretical models\cite{millis97,roder96} in which  a strong
coupling of the conduction electrons to a local Jahn-Teller distortion is considered to be
important for understanding the basic physical properties of manganites. 
In particular we found the following:\\
\noindent
1. The Mn ions are in the high-spin state in LaMnO$_3$ and hence the
contribution to the minority-spin channel from the Mn 3$d$ electrons are negligible small.\\

\noindent
2. Both Mn $t_{2g}$ and the $e_{g}$ electrons are present in the entire valence band region.
The $e_{g}$ electrons with the $dz^{2}$ symmetry are well localized compared with
those of $d_{x^{2}-y^{2}}$ symmetry and hence the $d_{z^{2}}$ contribution in the vicinity
of $E_{F}$ is negligible. On the other hand, the $t_{2g}$ electrons with $d_{xz}$ character
have large electron distributions on the top of the valence band comparable with the
distribution of $d_{x^{2}-y^{2}}$ electrons.\\

\noindent
3. The insulating behavior in LaMnO$_3$ originates from the combined effect of
Jahn-Teller distortion and A-type antiferromagnetic ordering.\\

\noindent
4. Without Jahn-Teller distortion LaMnO$_3$ is predicted to be a ferromagnetic metal.\\

\noindent
5. Unlike the LMTO-ASA calculations,\cite{hu00} without inclusion of the correlation effects
in the calculations we are able to predict the correct A-type antiferromagnetic insulating ground state for 
LaMnO$_3$ indicating that the importance of correlation effect has been exaggerated for 
this material.\\

\noindent
6. The large changes in energy and the development of an energy gap resulting from the
structural distortion along with exchange splitting indicate strong magnetostructural
coupling. This may be the possible origin for the observation of CMR in hole-doped LaMnO$_3$.\\

\noindent
7. Our calculation support the view point that the stabilization of the cubic phase is the
prime cause for the occurrence of the ferromagnetic state by hole doping rather than the often 
believed double-exchange mechanism.\\

\noindent
8. The density-functional approach works well for not only the ground-state properties 
but also the excited state properties of LaMnO$_3$.\\

\acknowledgements
PR is thankful for the financial support from the Research
Council of Norway. Also wish to acknowledge Y.Tokura, T.Arima, K.Takenaka
and J.Jung for generously sending there experimental optical data for comparison.
Part of these calculations were carried out on the
Norwegian supercomputer facilities. PR is grateful to Igor Solovyev and Lars Nordstr\"{o}m
for the useful communications and R. Vidya for the comments and critical reading of
this manuscript. John Wills, Karlheinz Schwarz and Peter Blaha for making their programs
available for the present study. O.E is grateful to the Swedish Research Foundation and the
Swedish Foundation for Strategic Research (SSF).

\begin{table}
\caption {
Total energy (relative to the lowest energy state in meV/f.u.) for LaMnO$_3$
in A-, C-, G-AF, F and P states with undistorted cubic (refered as cubic)
and orthorhombic structures. 
}
\begin{tabular}{l|c|c|c|c|c|l}
Method                &      A-AF  &      C-AF  &       G-AG    &     F &  P &   Reference  \\
\hline
LSDA(LAPW)            &      0   &     52   &       42   &     18  & ---     & \onlinecite{sawada97}\\
LSDA (LMTO-ASA, cubic) &     15   &    ---   &      125   &     --- &  ---    & \onlinecite{hu00}\\
LSDA(LMTO-ASA)        &     15   &    ---   &      110   &     0   &  ---    & \onlinecite{hu00}\\
LSDA+U (LMTO-ASA, cubic)&    0    &    ---   &     2042   &    1076 &  ---    & \onlinecite{hu00}\\
LSDA+U (LMTO-ASA)     &     0    &    ---   &      204   &     34  &  ---    & \onlinecite{hu00}\\
GGA+SO (FLMTO)        &     0    &    66.5  &     60.4   &    24.7 & 1515.4  & Present\\
GGA+SO (FLMTO, cubic)  &   406.9  &   439.4  &    573.1   &   347.5 & 1642.9  & Present\\
\end{tabular}
\label{table:de}
\end{table}
\begin{table}
\caption {Calculated magnetic moment (in $\mu_{B}/atom$) per Mn atom 
for LaMnO$_3$ in AF
(A-, C-, G-type) and F states. The
total moment (total) refers to the total magnetic moment per formula
unit.  
}
\begin{tabular}{l|c|c|c|c|c|l}
Method                 &  A    &    C   &       G    &      F &  F, total &  Reference\\
\hline
LSDA (LMTO-ASA, cubic)  & 3.81  &  ---   &     3.64   &    3.81 &    ---   &  \onlinecite{hu00}\\
LSDA (LMTO-ASA)        & 3.73  &  ---   &     3.63   &    3.76 &  ---   &  \onlinecite{hu00}\\
LSDA+U (LMTO-ASA)      & 3.72  & ---    &    3.63    &   3.76  & ---    & \onlinecite{hu00}\\
GGA+SO (FLMTO)         & 3.433 & 3.356  &    3.272   &   3.509 & 3.976  & Present\\
GGA+SO (FLMTO, cubic)   & 3.315 & 3.278  &    3.134   &   3.318 & 3.698  & Present\\
GGA    (FLAPW)         & 3.394 & ---    &    ---     &   3.295 & 3.525  & Present\\
Experiment             & 3.7$\pm$0.1&     &            &         &        & \onlinecite{elemans71}\\
\end{tabular}
\label{table:moment}
\end{table}
\begin{figure}
\caption{
The GdFeO$_3$-type crystal structure of LaMnO$_3$, viz. an orthorhombically
distorted perovskite-type structure.
}
\label{fig:str}
\end{figure}
\begin{figure}
\caption{Orbital projected DOS for Mn in F state
LaMnO$_3$ with hypothetical cubic perovskite structure. 
(Obtained by LAPW calculations.)}
\label{fig:cldos}
\end{figure}
\begin{figure}
\caption{Total DOS for LaMnO$_3$ in cubic perovskite-type atomic 
arrangement and
P, F, A-, C- and G-AF magnetic states.}
\label{fig:ctdos}
\end{figure}
\begin{figure}
\caption{Total density of states for LaMnO$_3$ in orthorhombic GdFeO$_3$-type
structure and P,
F, A-, C- and G-AF magnetic states.}
\label{fig:tdos}
\end{figure}
\begin{figure}
\caption{Site and angular-momentum projected DOS for A-AF state
LaMnO$_3$ in orthorhombic GdFeO$_3$-type structure.}
\label{fig:pdos}
\end{figure}
\begin{figure}
\caption{Orbital projected DOS for Mn in A-AF state
LaMnO$_3$ in orthorhombic GdFeO$_3$-type structure. (Results
obtained from LAPW calculations.)}
\label{fig:ldos}
\end{figure}
\begin{figure}
\caption{Magneto-optical Kerr-rotation, ellipticity, Faraday rotation
and ellipticity spectra obtained for F state LaMnO$_3$ in the 
orthorhombic GdFeO$_3$-type structure. The experimental Kerr-rotation spectrum for
La$_{0.7}$Sr$_{0.3}$MnO$_3$ 
is taken from Ref.\,\protect\onlinecite{popma75}
and the experimental Kerr-rotation 
and Kerr-ellipticity spectra for 
La$_{0.8}$Sr$_{0.2}$MnO$_3$ are taken from Ref.\,\protect\onlinecite{yamaguchi98}.}
\label{fig:moke}
\end{figure}
\begin{figure}
\caption{Imaginary part of optical dielectric tensors for the
diagonal elements ($\epsilon_{2}(\omega)$) for the A-AF states 
LaMnO$_3$ in hypothetical cubic and real orthorhombic GdFeO$_3$-type structures. Off-diagonal
element for the imaginary part of the dielectric tensor ($\epsilon_{2}^{xy}(\omega)$)
for F state LaMnO$_3$ with the GdFeO$_3$-type orthorhombic structure. The
experimental $\epsilon_{2}(omega)$ data are taken from Ref.\,\protect\onlinecite{jung97,arima93}
and the theoretical data are from Ref.\,\protect\onlinecite{solovyev96}.}
\label{fig:e2}
\end{figure}
\begin{figure}
\caption{Calculated electronic energy band structure for A-AF state LaMnO$_3$ in  
orthorhombic GdFeO$_3$ structure. Fermi level set at zero. Relevant symmetry
directions in the Brillouin zone are indicated.}
\label{fig:bnd}
\end{figure}
\begin{figure}
\caption{Calculated linear optical properties; absorption
coefficient (I($\omega$)) in 10$^{5}$ cm$^{-1}$), reflectivity (R($\omega$)), 
real part of optical dielectric tensor ($\epsilon_{1}(\omega$)), refractive
index ($n$($\omega$)), extinction coefficient ($k$($\omega$)), electron energy
loss function ($L$($\omega$)), and effective number of electrons participating the
optical interband transition ($N_{eff}$($\omega$) in electrons/f.u.)
for A-AF state LaMnO$_3$ in GdFeO$_3$-type
orthorhombic structure. The
experimental reflectivity spectra are taken from Ref.\,\protect\onlinecite{jung97,arima93,takenaka99}.} 
\label{fig:optic}
\end{figure}
\begin{figure}
\caption{XPS and BIS spectra for A-AF state LaMnO$_3$ in GdFeO$_3$-type 
orthorhombic structure. Experimental XPS spectrum is taken from
Ref.\,\protect\onlinecite{saitoh95} and BIS spectrum from Ref.\,\protect\onlinecite{sarma95}.
}
\label{fig:xps}
\end{figure}
\begin{figure}
\caption{XANES spectra for A-AF state LaMnO$_3$ in 
orthorhombic GdFeO$_3$-type structure. The experimental Mn $K$-edge spectra are taken
from Refs.\,\protect\onlinecite{subias98,bridges00,croft97}, the Mn $L_{II,III}$
spectrum from Ref.\,\protect\onlinecite{kurata93} and the O $K$-edge spectra
from Refs.\,\protect\onlinecite{park96,abbate99}.}
\label{fig:xanes}
\end{figure} 

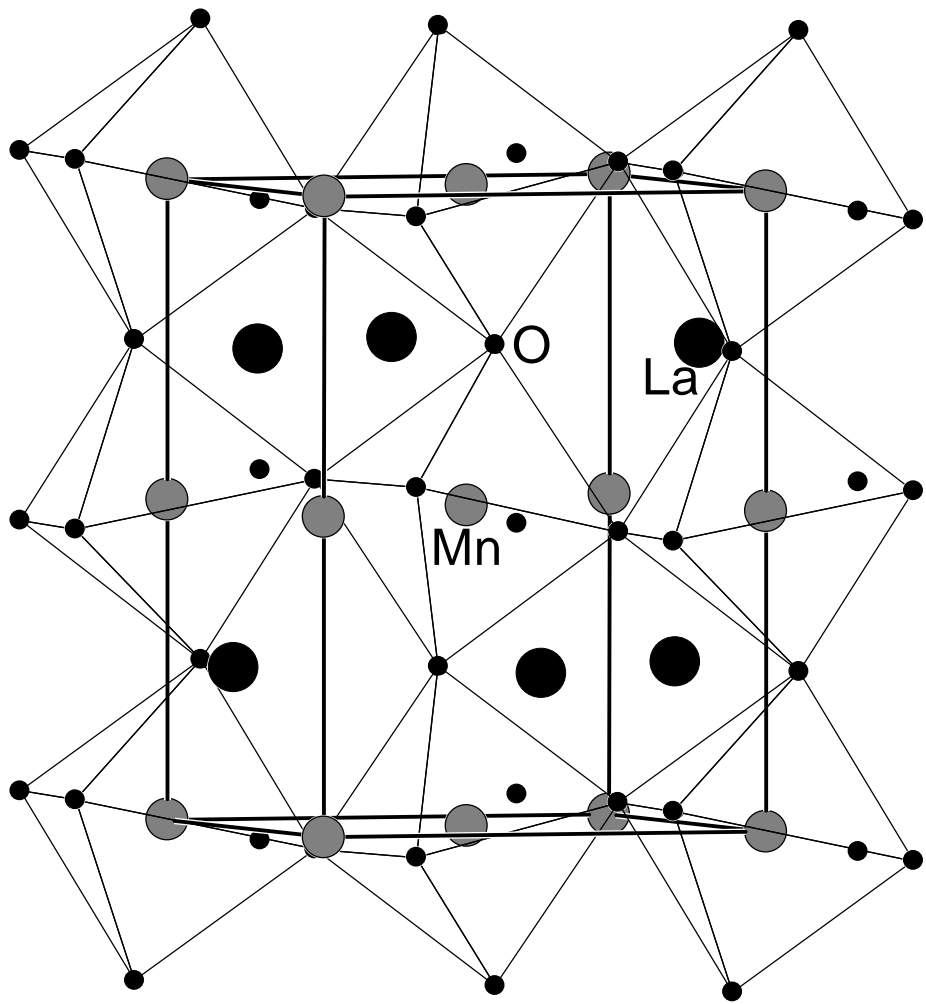
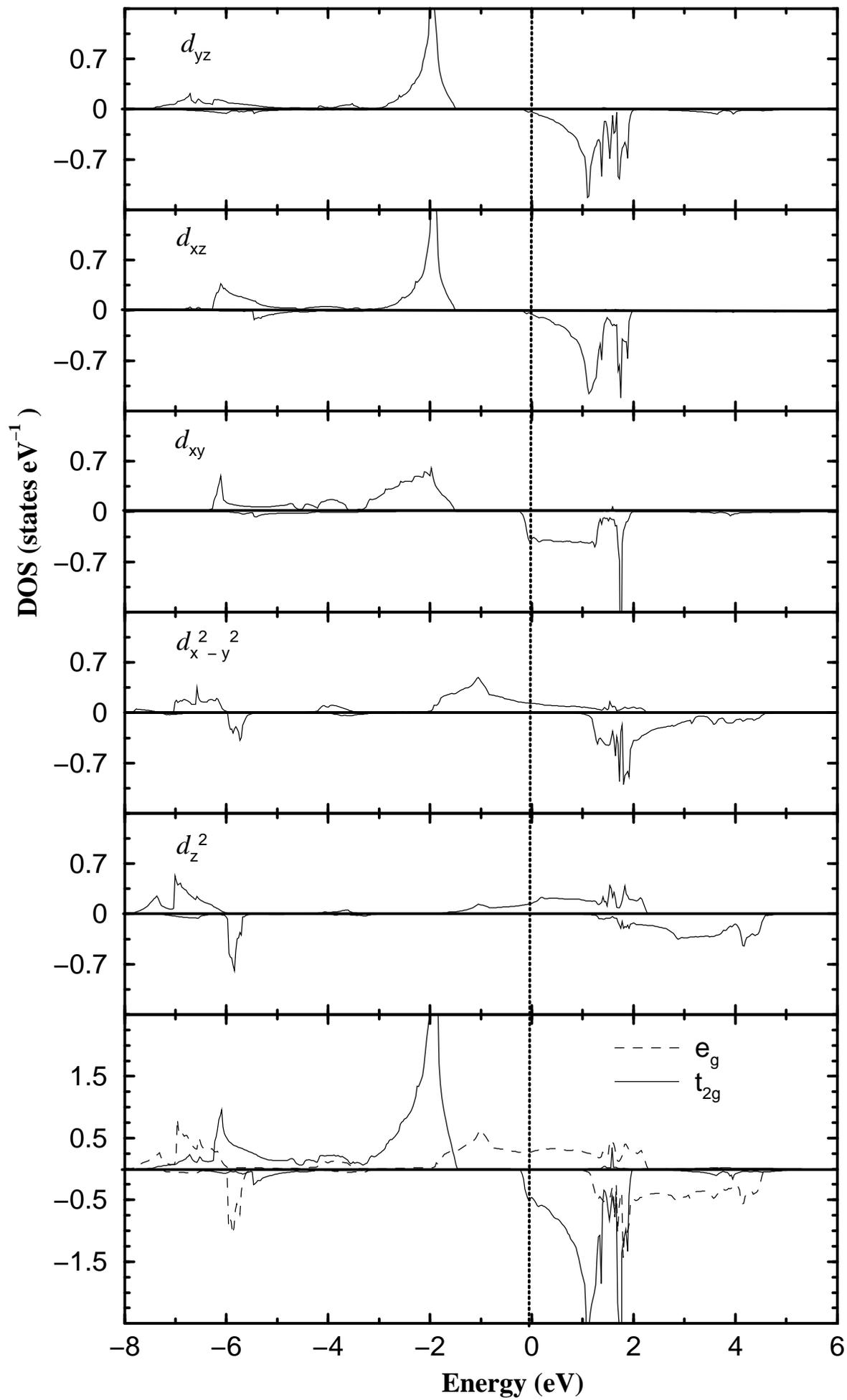
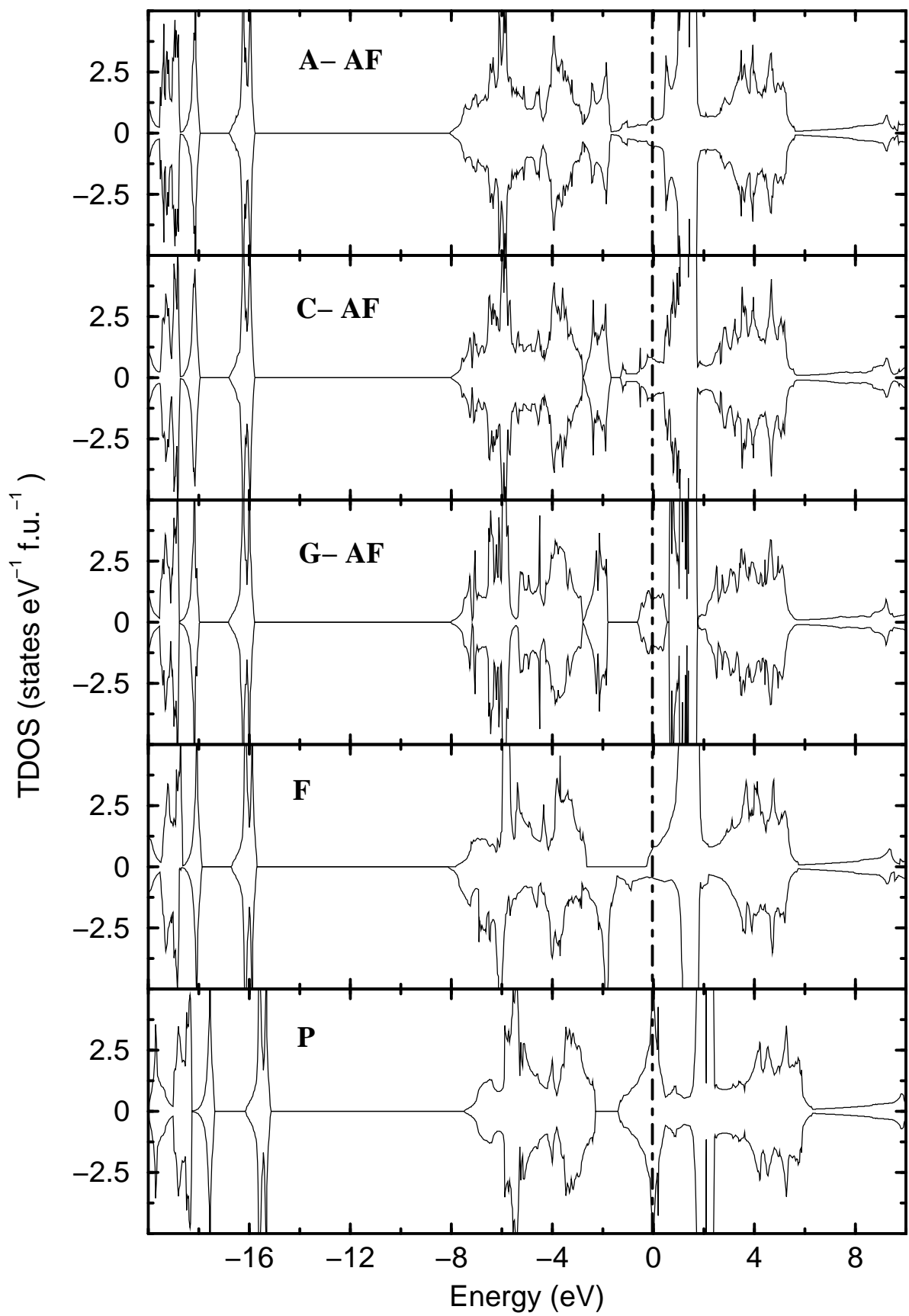
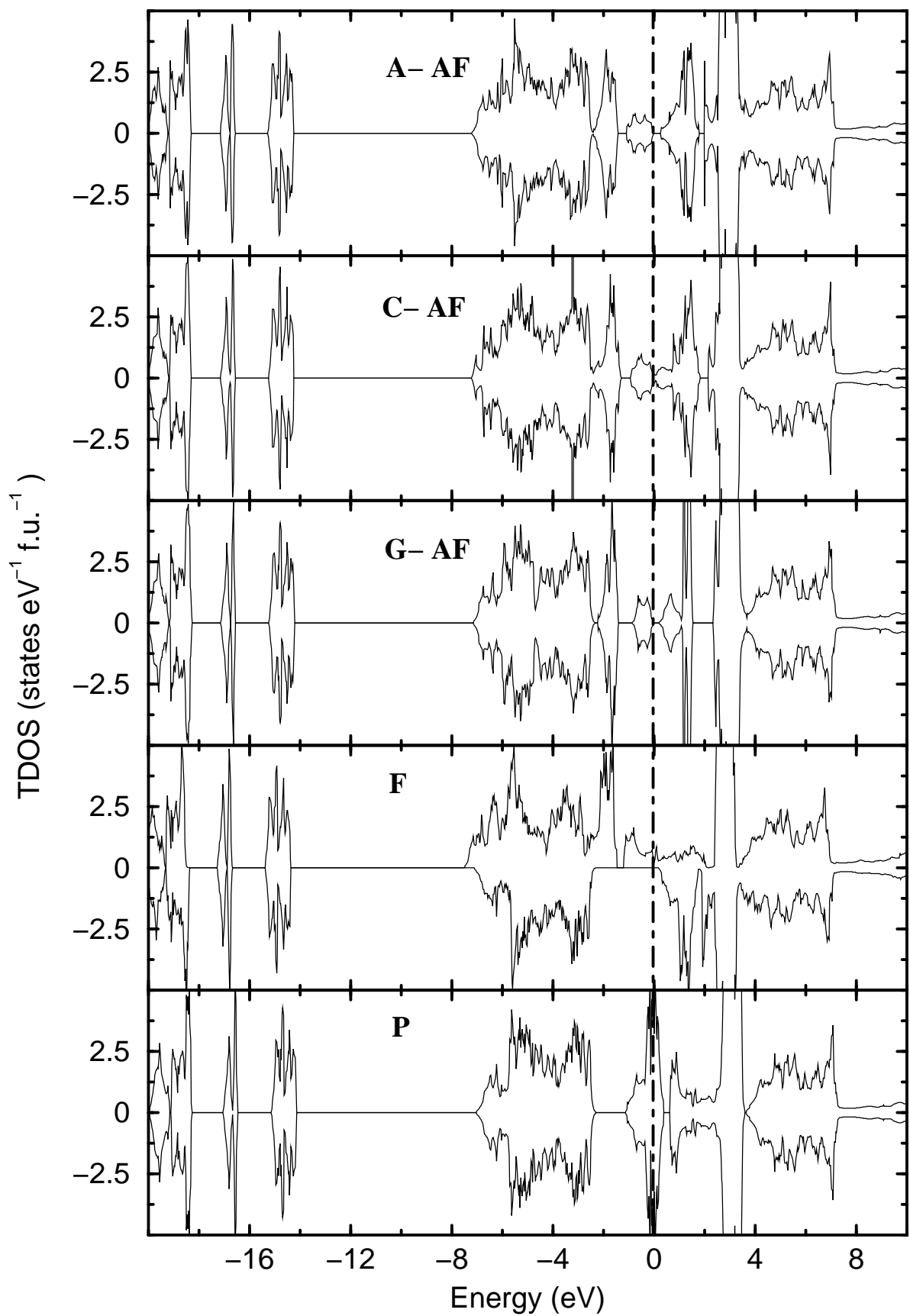
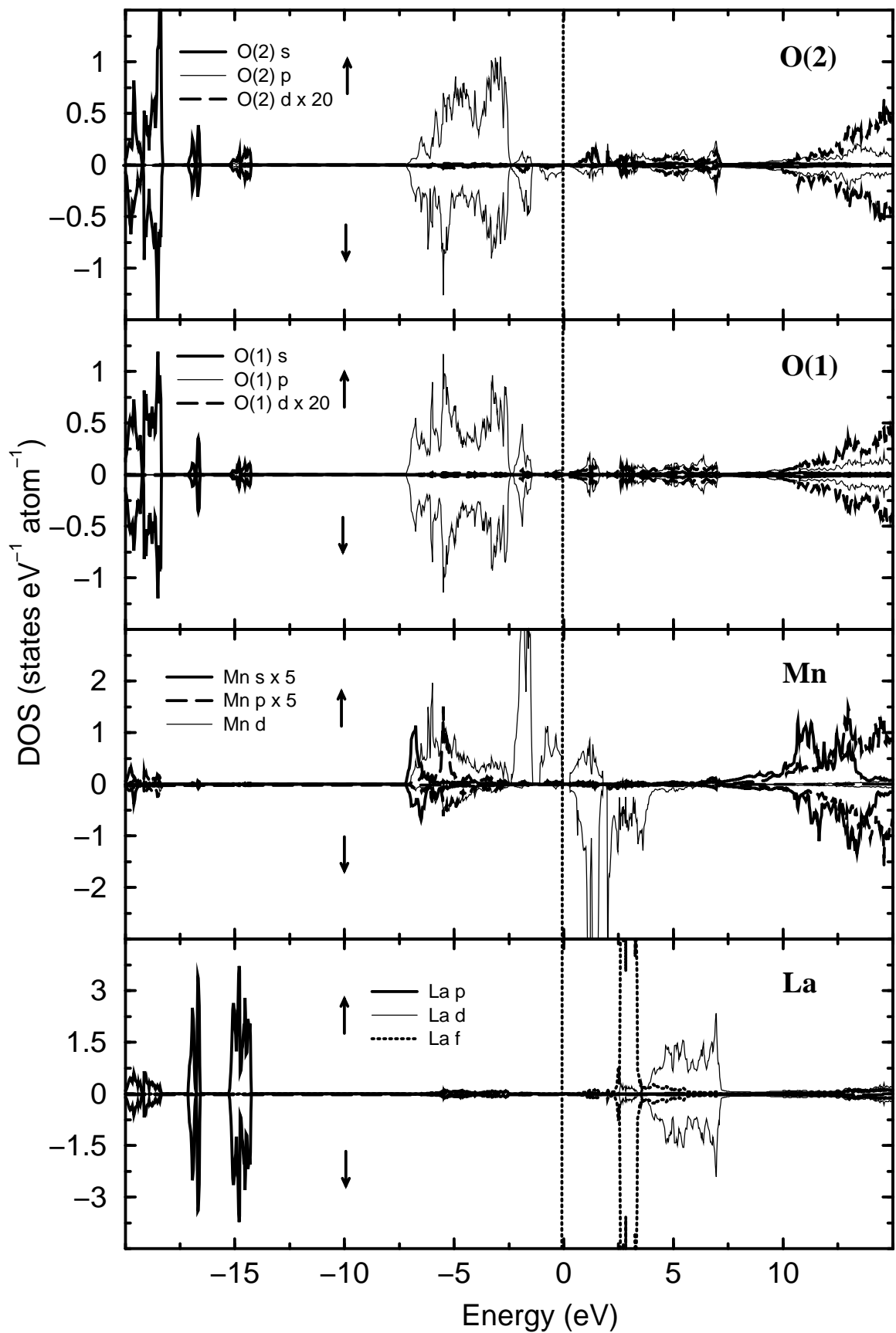
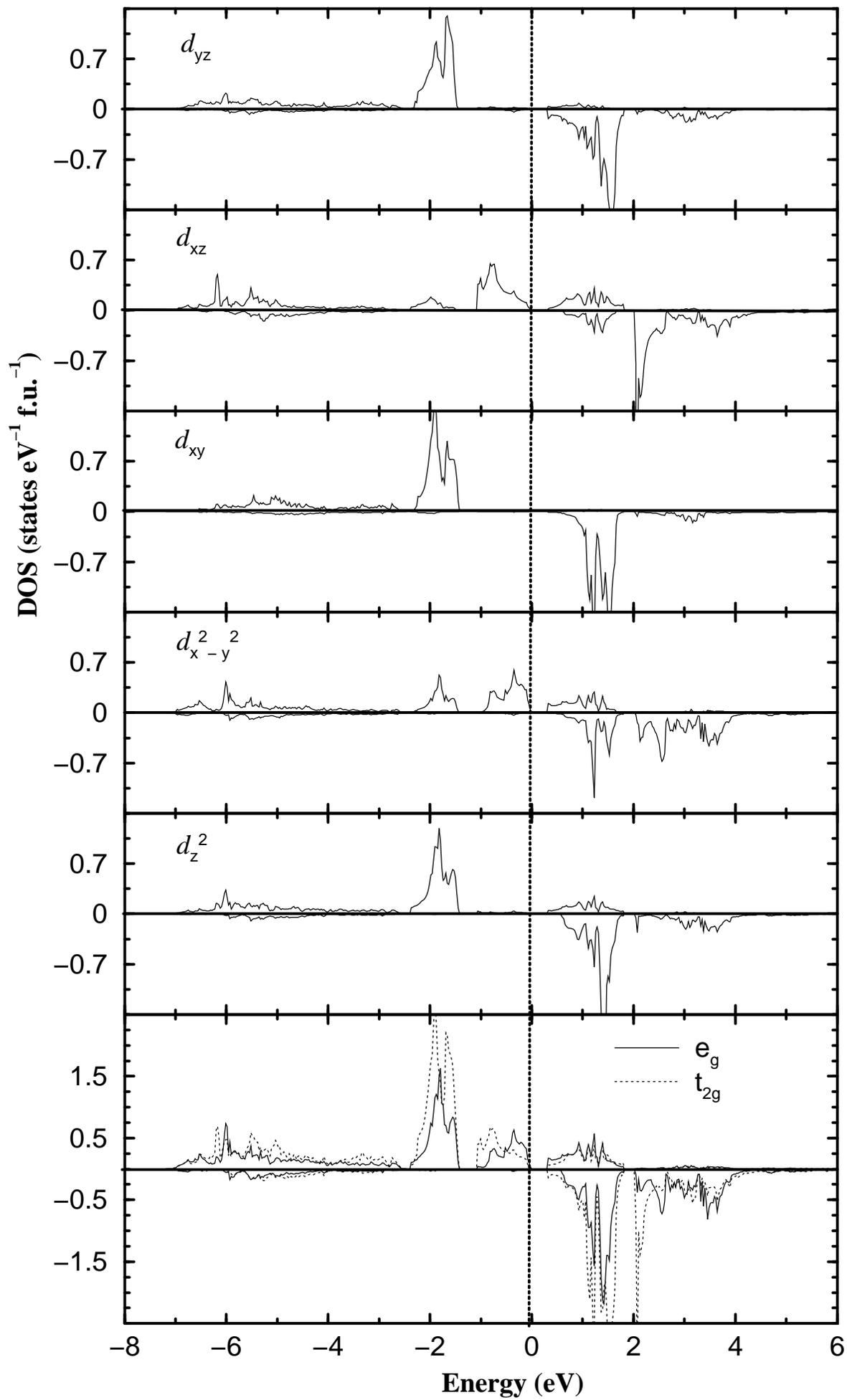
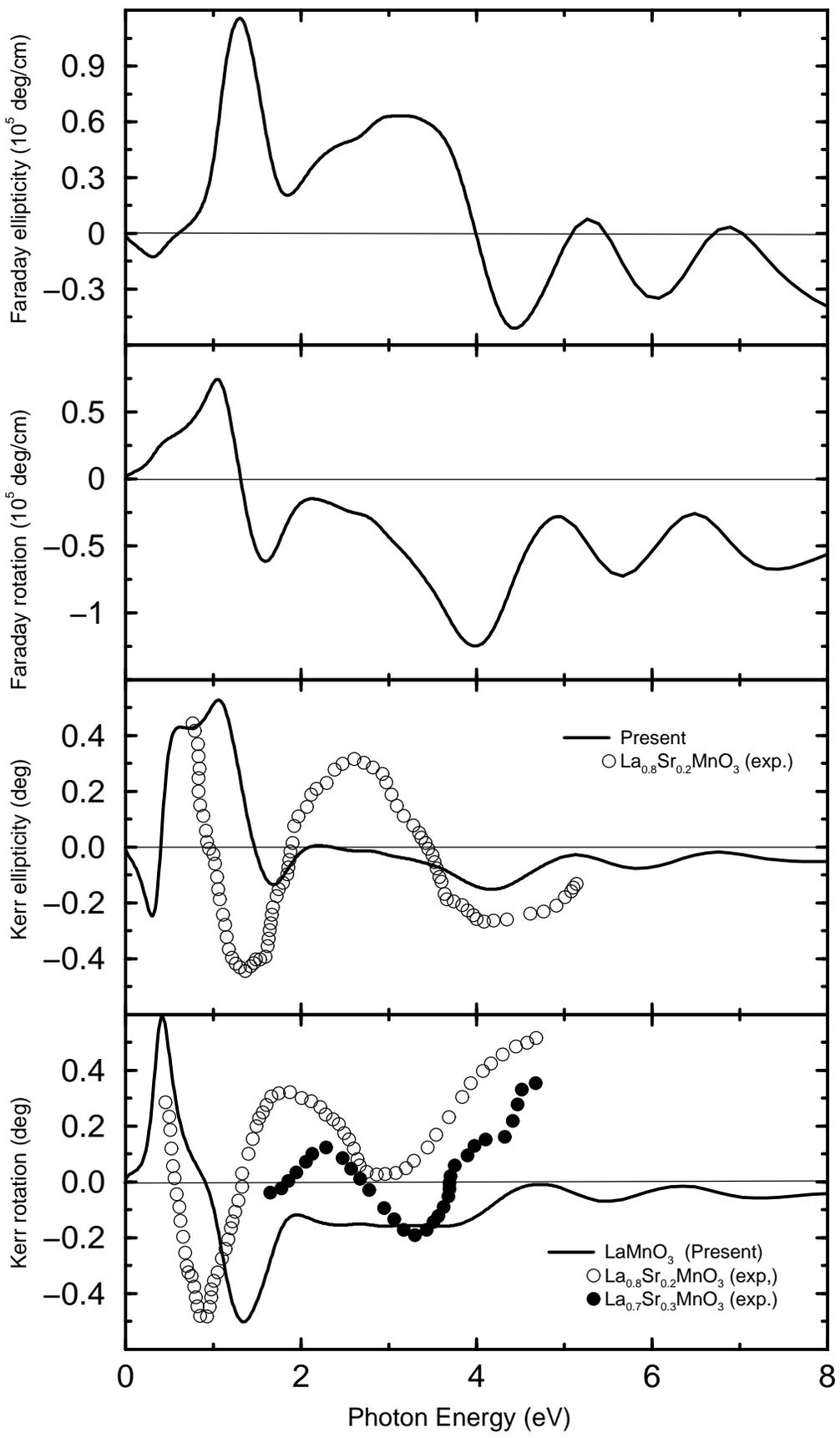
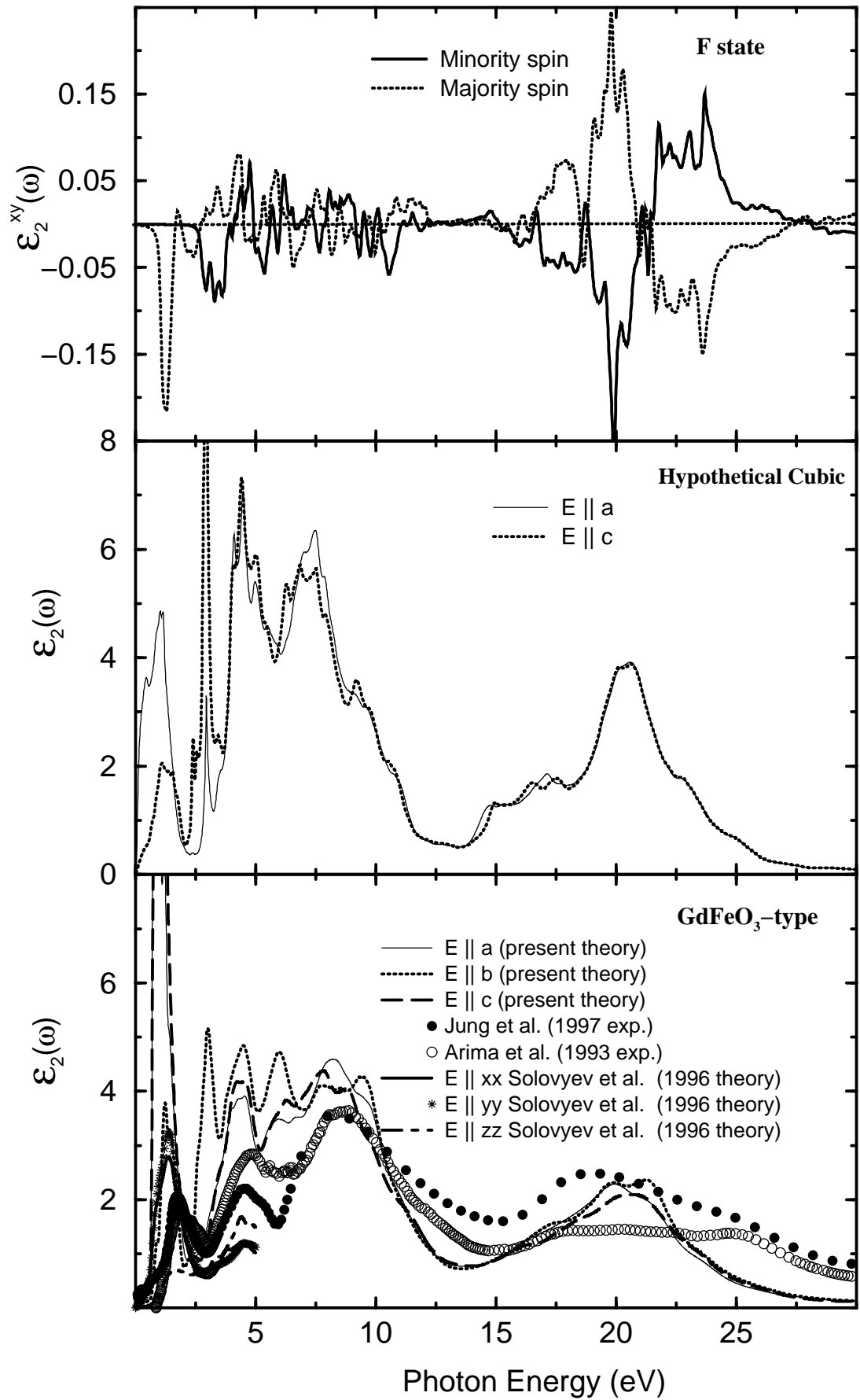
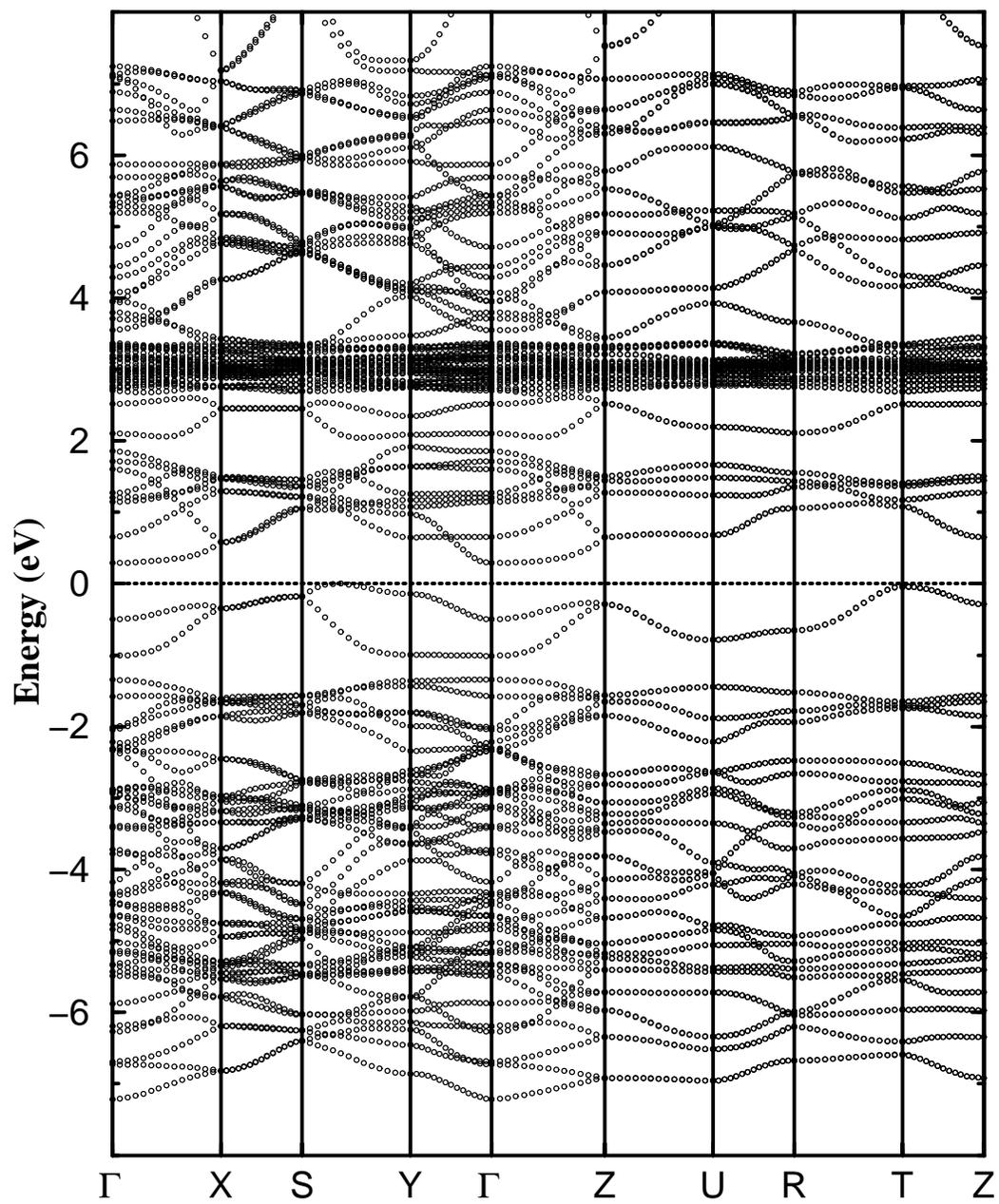
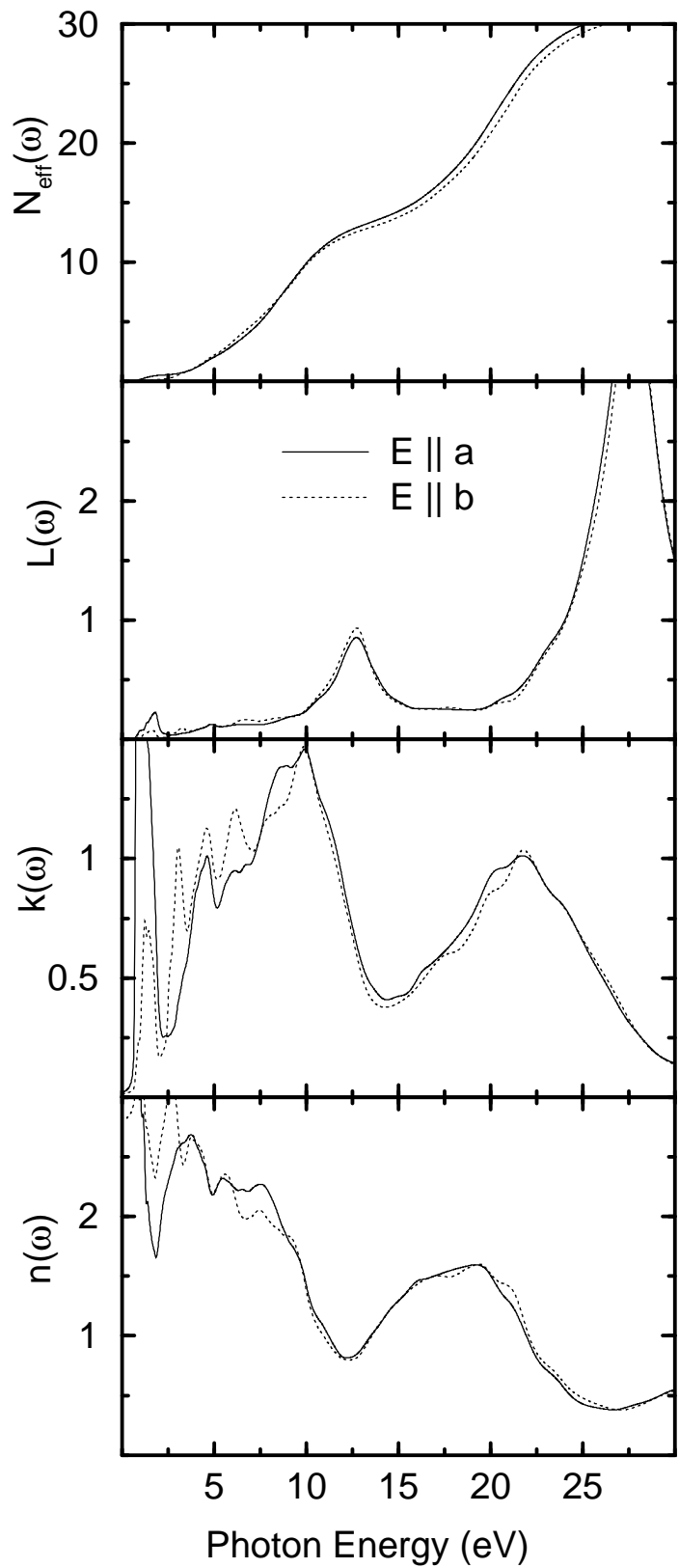
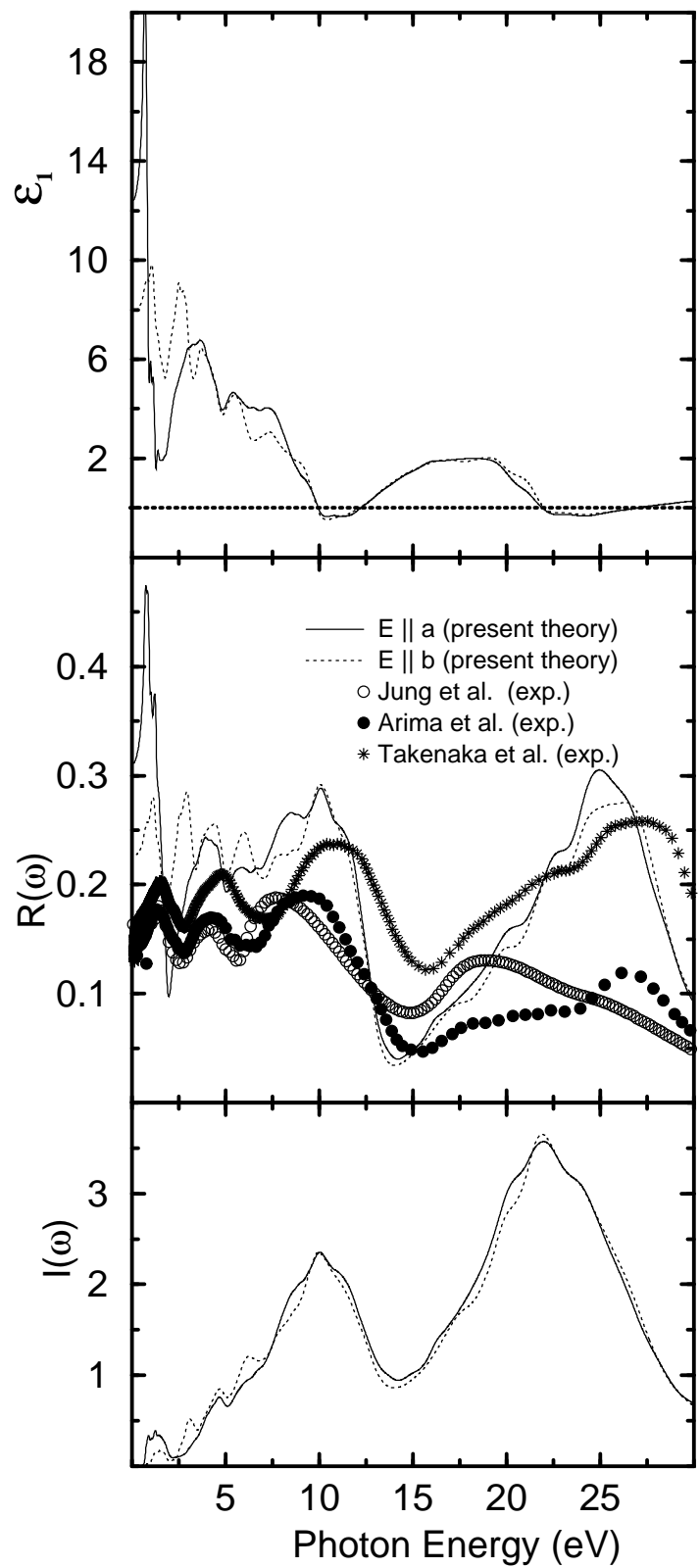
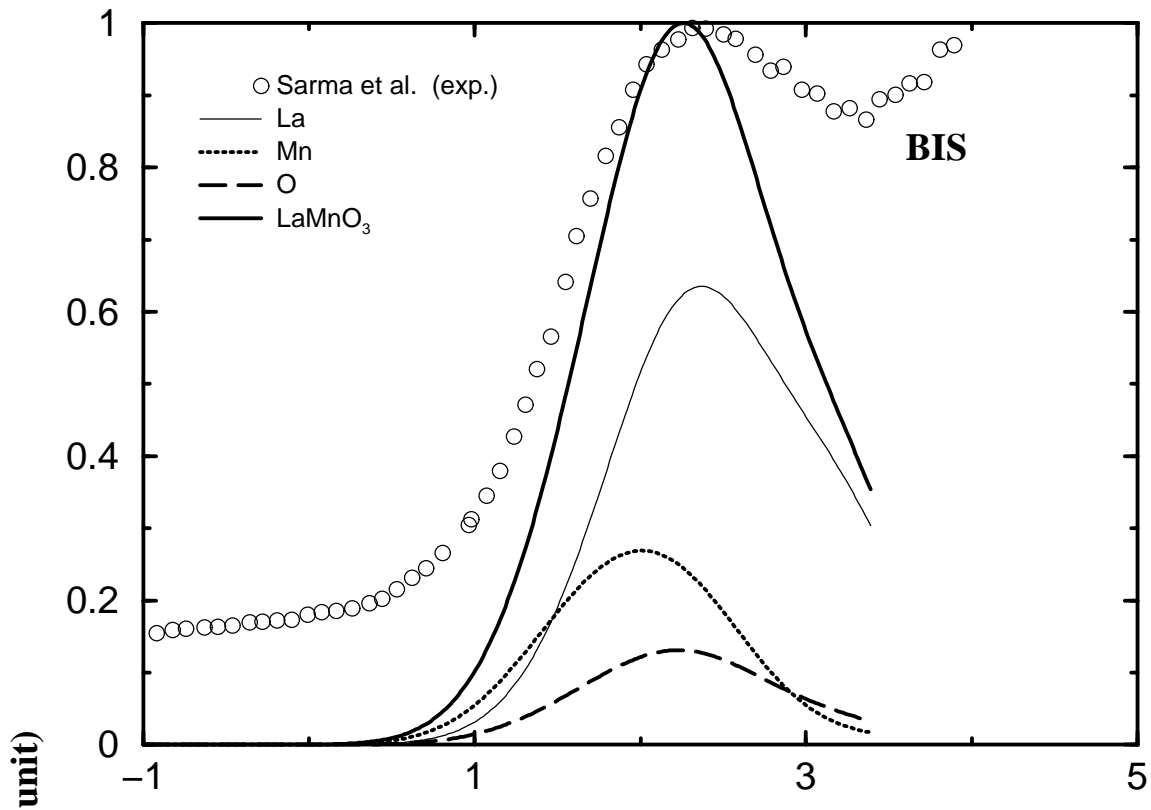
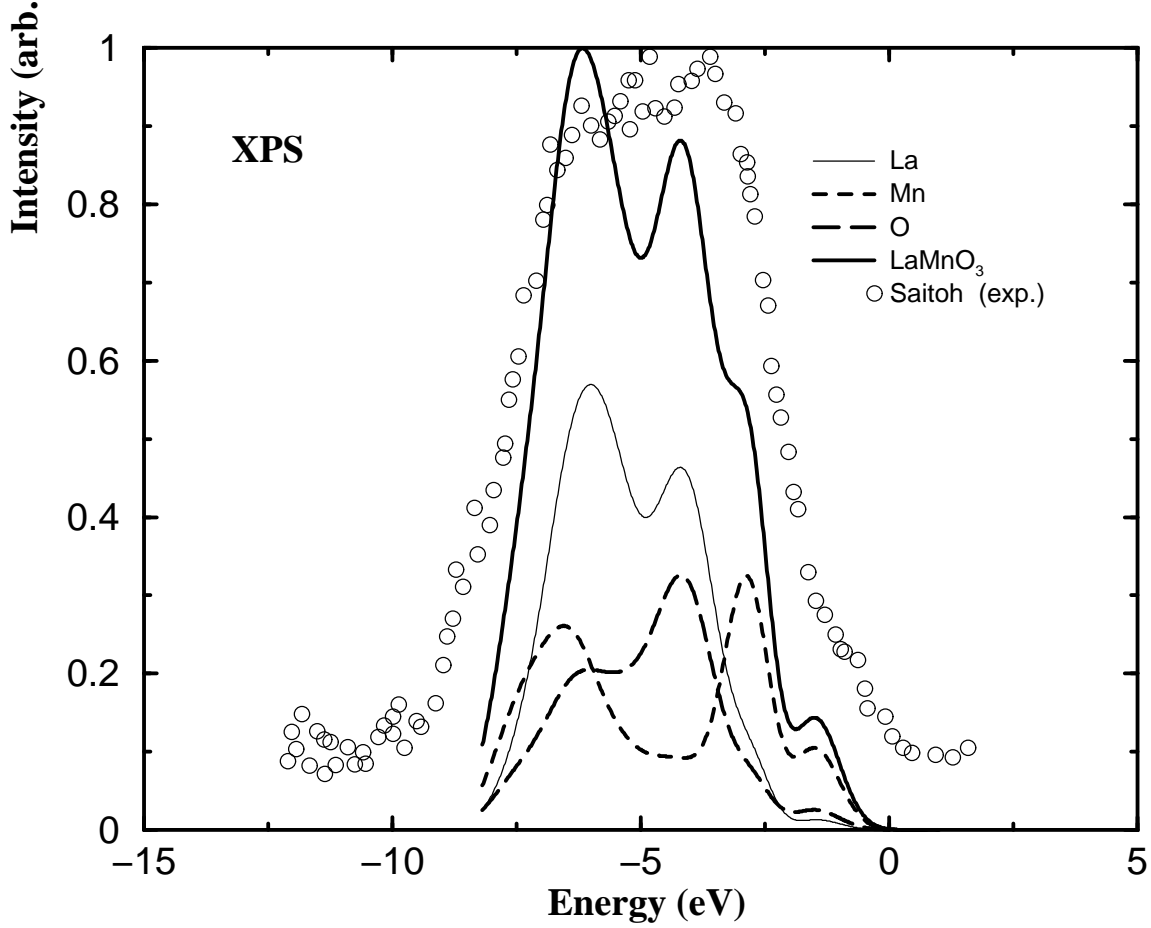
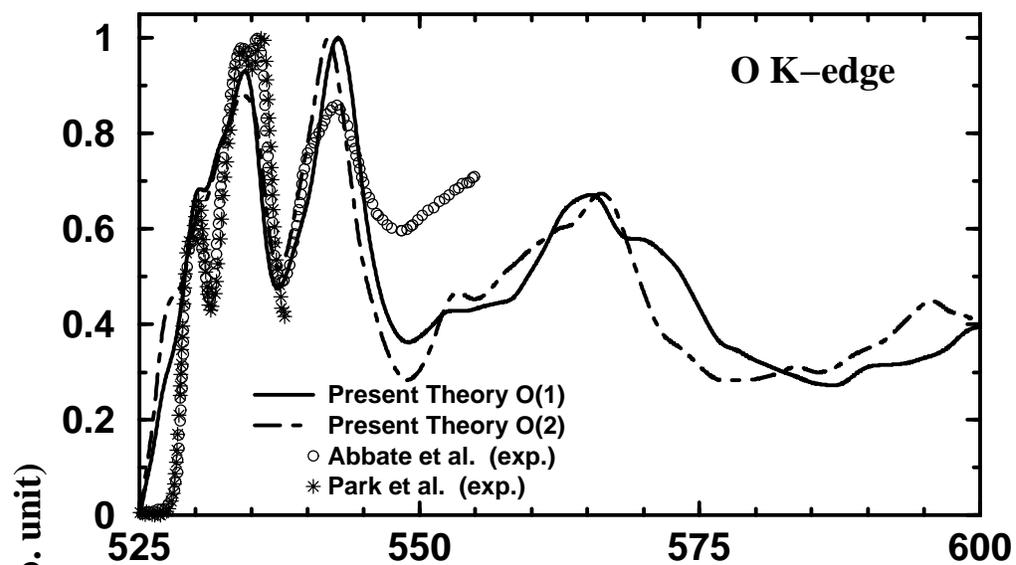
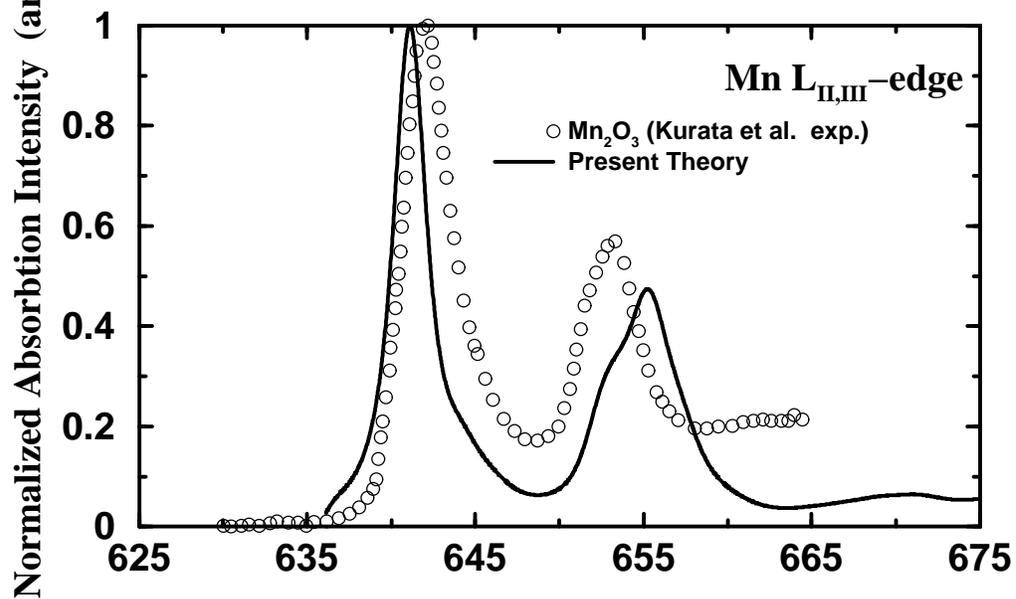
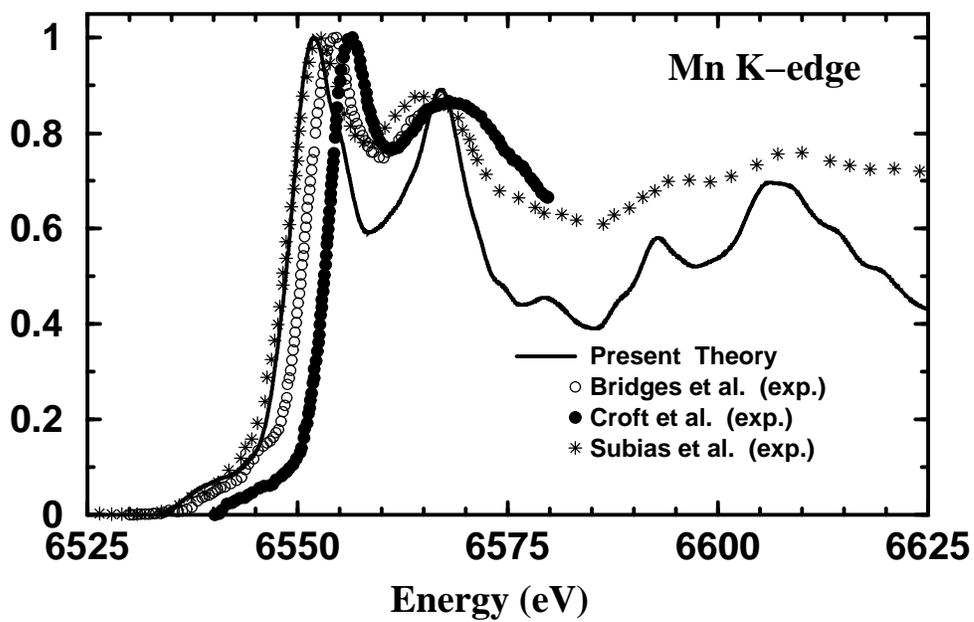
\end{document}